\documentclass[journal]{IEEEtran}

\usepackage{epsfig,amsmath,amssymb,amsfonts,amstext}
\usepackage{latexsym,graphics,epsf,epsfig,psfrag}

\newtheorem{definition}{\textbf{Definition}}
\newtheorem{corollary}{\textbf{Corollary}}
\newtheorem{lemma}{\textbf{Lemma}}
\newtheorem{theorem}{\textbf{Theorem}}
\newtheorem{proposition}{\textbf{Proposition}}
\newtheorem{remark}{\textbf{Remark}}

\newenvironment{myproof1}{{\em Proof of Convergenece:}}{\hfill$\bf
\blacksquare $\medskip}

\newcommand{\pp}{\hspace{1cm}}
\newcommand{\spp}{\hspace{5mm}}

\newcommand{\mE}{\mathrm{E}}

\newcommand{\cR}{\mathcal{R}}
\newcommand{\cX}{\mathcal{X}}

\newcommand{\cY}{\mathcal{Y}}

\newcommand{\cL}{\mathcal{L}}
\newcommand{\cW}{\mathcal{W}}

\newcommand{\cC}{\mathcal{C}}
\newcommand{\cN}{\mathcal{N}}

\newcommand{\cG}{\mathcal{G}}

\newcommand{\uP}{\underline{P}}
\newcommand{\ubeta}{\underline{\beta}}
\newcommand{\ut}{\underline{t}}

\newcommand{\uh}{\underline{h}}

\newcommand{\hw}{\hat{w}}

\newcommand{\barbeta}{\bar{\beta}}

\newcommand{\bartheta}{\bar{\theta}}
\begin{document}

\title{Resource Allocation for Wireless
Fading Relay Channels: Max-Min Solution}

\author{Yingbin~Liang,~\IEEEmembership{Member,~IEEE,}
Venugopal~V.~Veeravalli,~\IEEEmembership{Fellow,~IEEE,} H.~Vincent~Poor,
~\IEEEmembership{Fellow,~IEEE} % <-this % stops a space
\thanks{The material in
this paper was presented in part at the Asilomar Conference on
Signals, Systems and Computers, Pacific Grove, California, Nov.
2004.}
\thanks{This research was supported by the National Science Foundation under
CAREER/PECASE Grant CCR-00-49089 and Grants ANI-03-38807 and
CNS-06-25637, and by a Vodafone Foundation Graduate Fellowship.}
\thanks{Yingbin Liang is with the Department of Electrical Engineering, Princeton
University, Engineering Quadrangle, Olden Street, Princeton, NJ
08544; e-mail: {\tt yingbinl@princeton.edu}; Venugopal V.
Veeravalli is with the Department of Electrical and Computer
Engineering and the Coordinated Science Laboratory, University of
Illinois at Urbana-Champaign, 106 CSL, 1308 West Main Street,
Urbana, IL 61801; e-mail: {\tt vvv@uiuc.edu}; H. Vincent Poor is
with the Department of Electrical Engineering, Princeton
University, Engineering Quadrangle, Olden Street, Princeton, NJ
08544; e-mail: {\tt poor@princeton.edu} }}

%\markboth{Journal of \LaTeX\ Class Files,~Vol.~1, No.~11,~November~2002}
%{Shell \MakeLowercase{\textit{et al.}}: Bare Demo of IEEEtran.cls for Journals}

\maketitle

\begin{abstract}
Resource allocation is investigated for fading relay channels
under separate power constraints at the source and relay nodes. As
a basic information-theoretic model for fading relay channels, the
parallel relay channel is first studied, which consists of
multiple independent three-terminal relay channels as subchannels.
Lower and upper bounds on the capacity are derived, and are shown
to match, and thus establish the capacity for the parallel relay
channel with degraded subchannels. This capacity theorem is
further demonstrated via the Gaussian parallel relay channel with
degraded subchannels, for which the synchronized and asynchronized
capacities are obtained. The capacity achieving power allocation
at the source and relay nodes among the subchannels is partially
characterized for the synchronized case and fully characterized
for the asynchronized case. The fading relay channel is then
studied, which is based on the three-terminal relay channel with
each communication link being corrupted by a multiplicative fading
gain coefficient as well as an additive Gaussian noise term. For
each link, the fading state information is assumed to be known at
both the transmitter and the receiver. The source and relay nodes
are allowed to allocate their power adaptively according to the
instantaneous channel state information. The source and relay
nodes are assumed to be subject to separate power constraints. For
both the full-duplex and half-duplex cases, power allocations that
maximize the achievable rates are obtained. In the half-duplex
case, the power allocation needs to be jointly optimized with the
channel resource (time and bandwidth) allocation between the two
orthogonal channels over which the relay node transmits and
receives. Capacities are established for fading relay channels
that satisfy certain conditions.
\end{abstract}


\IEEEpeerreviewmaketitle

\section{Introduction}

The three-terminal relay channel was introduced by van der Meulen
\cite{Meul71} and was initially studied primarily in the context
of multiuser information theory \cite{Meul71,Cover79,Gamal82}. In
recent years, relaying has emerged as a powerful technique to
improve the reliability and throughput of wireless networks. An
understanding of wireless relay channels has thus become an
important area of research. Wireless relay channels and networks
have been addressed from various aspects, including
information-theoretic capacity
\cite{Send03I,Send03II,Schein00,Xie04,Kramer05,Host05,Host06,Wang05,
Katz05,Gastpar05,Mitra03,Khoj04aller,Nabar04,Sankar04_1,Jindal04,Liang05itb,Dabora05,Gamal06},
diversity \cite{Laneman03,Azarian05,Prasad04,Yuksel06}, outage
performance \cite{Gunduz06,Aves06}, and cooperative coding
\cite{Stef04,Murugan06,Bao06}. Central to the study of wireless
relay channels is the problem of resource allocation. For example,
the source and relay nodes can dynamically allocate their transmit
powers to achieve a better rate if the fading state information is
available. Resource allocation for relay channels and networks has
been studied by several recent papers, including
\cite{Host05,Yao05,Reznik04,Maric04isit,Maric04asilomar,Gunduz06}.
Common to all of these studies is the assumption that the source
and relay nodes are subject to a total power constraint.

In this paper, we study wireless fading relay channels, where we
assume that the source and relay nodes are subject to separate
power constraints instead of a total power constraint. This
assumption is more practical for wireless networks, because the
source and relay nodes are usually geographically separated, and
are hence supported by separate power supplies. Under this
assumption, the resource allocation problem falls under a class of
{\it max-min} problems. We connect such {\it max-min} problems to
the {\it minimax} two hypothesis testing problem (see, e.g.,
 \cite[II.C]{Poor94}), and apply a similar technique to find
optimal (in the {\it max-min} sense) resource allocation
strategies for fading relay channels.

We first study the parallel relay channel, which consists of
multiple independent relay channels and serves as a basic
information-theoretic model for fading relay channels. We derive a
lower bound on the capacity based on the partial
decode-and-forward scheme as well as a cut-set upper bound. We
show that the two bounds match and establish the capacity for the
parallel relay channel with degraded subchannels. This generalizes
the capacity result in \cite[Th.~12]{Liang0606it} to multiple
subchannels. We also demonstrate that the parallel relay channel
is not a simple combination of subchannels in that the capacity of
the parallel relay channel can be larger than the sum of the
capacities of subchannels, as was also remarked in
\cite[Sec.~VII]{Liang0606it}.

We then study the Gaussian parallel relay channel with degraded
subchannels. There are two types of capacity that can be defined
for this channel. The first is the synchronized capacity, where
the source and relay inputs are allowed to be correlated. To
achieve the capacity, the source and relay nodes need to choose an
optimal correlation parameter for each subchannel, and further to
choose an optimal power allocation across the subchannels under
separate power constraints. We characterize the optimal solutions
for the cases where the optimization is convex, and provide
equations that need to be solved numerically for cases where the
optimization is nonconvex. We also study the asynchronized
capacity, where the source and relay inputs are required to be
independent. This capacity is easier to achieve in practice due to
the simpler transceiver design for the source and relay nodes. We
fully characterize the capacity-achieving power allocation at the
source and relay nodes in closed form.

We then move on to study the fading relay channel, which is based
on the classical relay channel with each transmission link being
corrupted by a multiplicative stationary and ergodic fading
process as well as an additive white Gaussian noise process. The
fading relay channel is a special case of the parallel relay
channel, with each subchannel corresponding to one fading state
realization. We assume that both the transmitter and the receiver
know the channel state information, so that the source and relay
nodes can allocate their transmission powers adaptively according
to the instantaneous fading state information. We consider the
resource allocation problem for two fading relay models:
full-duplex and half-duplex.

The fading full-duplex relay channel has been studied in
\cite{Host05}, where lower and upper bounds on the capacity were
derived, along with the resource allocation that optimizes these
bounds, under a total power constraint for the source and relay
nodes. In this paper, we assume separate power constraints for the
source and relay nodes and study the power allocation that
optimizes the capacity bounds. We focus on the more practical
asynchronized case. We obtain the power allocation that maximizes
an achievable rate, and show that the optimal power allocation may
be {\em two-level water-filling}, {\em orthogonal division
water-filling}, or {\em iterative water-filling} depending on the
channel statistics and the power constraints. We also establish
the asynchronized capacity for channels that satisfy a certain
condition.

We further study a fading half-duplex relay channel model, where
the source node transmits to the relay and destination nodes in
one channel, and the relay node transmits to the destination node
in an orthogonal channel. We introduce a parameter $\theta$ to
represent the channel resource (time and bandwidth) allocation
between the two orthogonal channels. We study three scenarios. In
Scenario I, where the two orthogonal channels share the channel
resource equally, i.e., $\theta=1/2$, we show that the optimal
power allocation falls into three cases depending on the ranges of
power constraints at the source and relay nodes. The optimal power
allocation for the relay node is always water-filling, but the
power allocation for the source node is not water-filling in
general. In scenario II, the channel resource allocation parameter
$\theta$ needs to be same for all channel states but can be
jointly optimized with the power allocation. In Scenario III,
which is the most general scenario, $\theta$ can change with
channel realizations and is jointly optimized with power
allocation. For both Scenarios II and III, we derive the jointly
optimal $\theta$ and power allocation that maximize the achievable
rate. Furthermore, we show that the lower bound achieves the
cut-set upper bound if the channel statistics and power constraint
satisfy a certain condition. We hence establish the capacity for
these channels over all possible power and channel resource
allocations.

The paper is organized as follows. In Section \ref{sec:pararelay},
the parallel relay channel is introduced and studied. In Section
\ref{sec:gauss}, the optimal resource allocation that achieves the
capacity for the Gaussian parallel relay channel with degraded
subchannels is studied. In Section \ref{sec:fullrelay}, resource
allocation for the fading full-duplex relay channel is presented.
In Section \ref{sec:halfrelay}, resource allocation for the fading
half-duplex relay channel is studied, where the three scenarios
described above are considered. Finally in Section
\ref{sec:conclusion}, we give concluding remarks.
\section{Parallel Relay Channels}\label{sec:pararelay}

\begin{figure*}[t]
\begin{center}
\includegraphics[width=13cm]{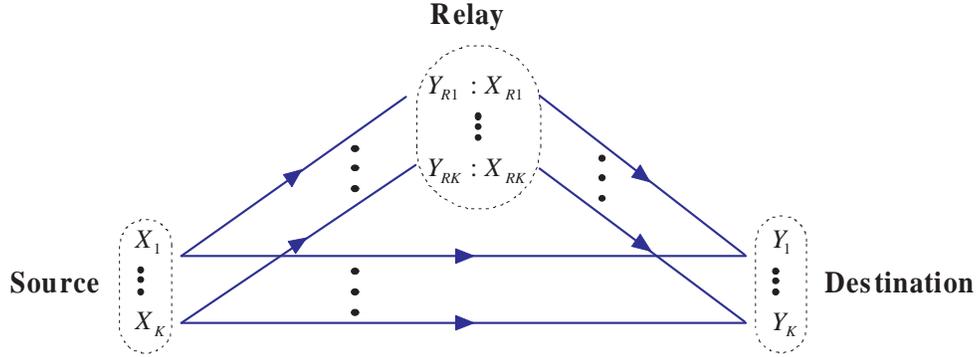}
\caption{Parallel relay channel} \label{fig:pararelay}
\end{center}
\end{figure*}

In this section, we study the parallel relay channel, which serves
as a basic information-theoretic model for the fading relay
channels that are considered in Sections \ref{sec:fullrelay} and
\ref{sec:halfrelay}. The parallel relay channel also models the
relay channel where the source and relay nodes can transmit over
multiple frequency bands with each subchannel corresponding to the
channel over one frequency band. It is shown in this section that
in contrast to the parallel point-to-point channel, the parallel
relay channel is not a simple combination of independent
subchannels.
\begin{definition}
A parallel relay channel with $K$ subchannels (see
Fig.~\ref{fig:pararelay}) consists of $K$ finite source input
alphabets $\cX_1,\ldots,\cX_K$, $K$ finite relay input alphabets
$\cX_{R1},\ldots,\cX_{RK}$, $K$ finite destination output
alphabets $\cY_1,\ldots, \cY_K$ and $K$ finite relay output
alphabets $\cY_{R1},\ldots,\cY_{RK}$. The transition probability
distribution is given by
\begin{equation}
\prod_{k=1}^K p_k(y_k,y_{Rk}| x_k,x_{Rk})
\end{equation}
where $x_k \in \cX_k$, $x_{Rk} \in \cX_{Rk}$, $y_k \in \cY_k$, and
$y_{Rk} \in \cY_{Rk}$ for $k=1,\ldots,K$.
\end{definition}

A $\left( 2^{nR}, n \right)$ code consists of the following:
\begin{list}{$\bullet$}{\topsep=0ex \leftmargin=5mm
\rightmargin=0mm \itemsep=0mm}

\item One message set $\cW=\{1,2,\ldots,2^{nR}\}$ with the message
$W$ uniformly distributed over $\cW$;

\item One encoder at the source node that maps each message $w \in
\cW $ to a codeword
\[
(x_{11},\ldots,x_{1n},\ldots,x_{K1},\ldots,x_{Kn});
\]

\item A set of relay functions $\left\{f_i\right\}_{i=1}^{n}$ such
that for $1 \leq i \leq n$:
\begin{equation*}
\begin{split}
&
(x_{R1i},\ldots,x_{RKi})\\
& \hspace{3mm}
=f_i(y_{R11},\ldots,y_{R1[i-1]},\ldots,y_{RK1},\ldots,y_{RK[i-1]});
\end{split}
\end{equation*}

\item One decoder at the destination node that maps a received
sequence $(y_{11},\ldots,y_{1n},\ldots,$ $ y_{K1},\ldots,y_{Kn}) $
to a message $\hw \in \cW$.
\end{list}

Note that the relay node is allowed to jointly encode and decode
across the K parallel subchannels.

A rate $R$ is {\em achievable} if there exists a sequence of
$\left(2^{nR}, n \right)$ codes with the average probability of
error at the destination node going to zero as $n$ goes to
infinity.

The following theorem provides lower and upper bounds on the
capacity of the parallel relay channel.
\begin{theorem}\label{th:paralowup}
For the parallel relay channel, a lower bound on the capacity is
given by
\begin{equation}\label{eq:paralow}
\begin{split}
C_{\mathtt{low}}=& \max\min \Bigg\{\sum_{k=1}^K I(X_k,X_{Rk};Y_k),
\\
& \sum_{k=1}^K I(Q_k;Y_{Rk}|X_{Rk})+I(X_k;Y_k|Q_k,X_{Rk}) \Bigg\}
\end{split}
\end{equation}
where $Q_k$ for $k=1,\ldots,K$ are auxiliary random variables. The
maximum in \eqref{eq:paralow} is over the joint distribution
\[\prod_{k=1}^K
p_k(q_k,x_{Rk},x_k)p_k(y_k,y_{Rk}| x_k,x_{Rk}).\]

An upper bound on the capacity is given by
\begin{equation}\label{eq:paraup}
\begin{split}
C_{\mathtt{up}}=\max \min \Bigg\{ & \sum_{k=1}^K
I(X_k,X_{Rk};Y_k),
\\
&\sum_{k=1}^K I(X_k;Y_k,Y_{Rk}|X_{Rk}) \Bigg\}
\end{split}
\end{equation}
where the maximum in \eqref{eq:paraup} is over the joint
distribution
\[\prod_{k=1}^K
p_k(x_{Rk},x_k)p_k(y_k,y_{Rk}| x_k,x_{Rk}).\]
\end{theorem}
\begin{remark}
The lower bound \eqref{eq:paralow} generalizes the rate given in
\cite[Theorem 1]{Liang05it} based on the decode-and-forward
scheme.
\end{remark}
\begin{proof}
To derive the lower bound \eqref{eq:paralow}, we use the following
achievable rate for the relay channel based on the partial
decode-and-forward scheme given in \cite{Gamal82}:
\begin{equation}
\begin{split}
R < & \max \min \\
&\Big\{ I(X_R,X;Y), \; I(Q;Y_R|X_R)+ I(X;Y|Q,X_R) \Big\}
\end{split}
\end{equation}
We set $Q=(Q_1,\ldots,Q_K)$, $X=(X_1,\ldots,X_K)$,
$X_R=(X_{R1},\ldots,X_{RK})$, $Y=(Y_1,\ldots,Y_K)$, and
$Y_R=(Y_{R1},\ldots,Y_{RK})$ in the above achievable rate. We
further choose $(Q_1,X_1,X_{R1})$, $\ldots$, $(Q_K,X_K,X_{RK})$ to
be independent, and then obtain the lower bound
\eqref{eq:paralow}.

The upper bound \eqref{eq:paraup} is based on the cut-set bound
\cite[Theorem 4]{Cover79} and the independency of the $K$ parallel
subchannels.
\end{proof}
\begin{remark}\label{rk:crossflow}
In the achievable scheme, the relay node first decodes information
sent by the source node over each subchannel. The relay node then
reassigns total decoded information to each subchannel to forward
to the destination node. Hence information that was sent to the
relay node over one subchannel may be forwarded to the destination
node over other subchannels, as long as the total rate at which
the relay node can forward information to the destination node
over all subchannels is larger than the total rate at which the
relay node can decode information from the source node.
\end{remark}

The lower and upper bounds in Theorem \ref{th:paralowup} do not
match in general. We next study a class of parallel relay channels
with degraded subchannels. For this channel, the lower and upper
bounds match, and we hence establish the capacity. Moreover, this
capacity provides an achievable rate for the case where the
subchannels are either stochastically degraded or reversely
degraded (e.g., fading relay channels).

\begin{definition}\label{def:pararelay}
Consider the parallel relay channel with degraded subchannels.
Assume each subchannel is either degraded or reversely degraded,
i.e., each subchannel satisfies either
\begin{equation}\label{eq:dg1}
\begin{split}
& p_k(y_k,y_{Rk}|x_k,x_{Rk})
\\
& \spp =p_k(y_{Rk}|x_k,x_{Rk})p_k(y_k|y_{Rk},x_{Rk}),
\end{split}
\end{equation}
or
\begin{equation}\label{eq:dg2}
\begin{split}
& p_k(y_k,y_{Rk}|x_k,x_{Rk}) \\
& \spp =p_k(y_k|x_k,x_{Rk})p_k(y_{Rk}|y_k,x_{Rk}).
\end{split}
\end{equation}
\end{definition}

We note that the parallel relay channel with degraded subchannels
has been studied in \cite[Sec.~VII]{Liang0606it} for the
two-subchannel case. We now generalize the result in
\cite[Sec.~VII]{Liang0606it} to channels with multiple
subchannels. In fact, our main focus is on the Gaussian case
considered in this section and Section \ref{sec:gauss}.

We define the set $A$ to contain the indices of the subchannels
that satisfy \eqref{eq:dg1}, i.e., those subchannels where the
source-to-relay channel is stronger than the source-to-destination
channel. Then the set $A^c$ contains the indices of the
subchannels that satisfy \eqref{eq:dg2}, i.e., those subchannels
where the source-to-relay channel is weaker than the
source-to-destination channel. Note that in general the parallel
relay channel with degraded subchannels is neither a degraded
relay channel nor a reversely degraded channel. For this channel,
the lower and upper bounds given in Theorem \ref{th:paralowup}
match and establish the following capacity theorem.
\begin{theorem}\label{th:dgcapa}
For the parallel relay channel with degraded subchannels, the
capacity is given by
\begin{equation}\label{eq:dgcapa}
\begin{split}
C=& \max \min \Bigg\{ \sum_{k=1}^K I(X_k,X_{Rk};Y_k), \\
& \sum_{k\in A} I(X_k;Y_{Rk}|X_{Rk})+\sum_{k\in
A^c}I(X_k;Y_k|X_{Rk}) \Bigg\}.
\end{split}
\end{equation}
where the maximum is over the joint distribution
\[\prod_{k=1}^K
p_k(x_{Rk},x_k)p_k(y_k,y_{Rk}| x_k,x_{Rk}).\]
\end{theorem}
\begin{remark}
Theorem \ref{th:dgcapa} generalizes the capacity of the parallel
relay channel with unmatched degraded subchannels in
\cite[Theorem~12]{Liang0606it} to channels with multiple
subchannels.
\end{remark}
\begin{proof}
The achievability follows from $C_{\mathtt{low}}$ in
\eqref{eq:paralow} by setting $Q_k=X_k$ for $k \in A$ and setting
$Q_k=\phi$ for $k \in A^c$. The converse follows from
$C_{\mathtt{up}}$ in \eqref{eq:paraup} by applying the
degradedness conditions \eqref{eq:dg1} and \eqref{eq:dg2}.
\end{proof}

Note that the partial decode-and-forward scheme achieves the
capacity of the parallel relay channel with degraded subchannels.
From the selection of $Q_k$ in the above achievability proof, it
can be seen that the relay node decodes all the information sent
over the degraded subchannels, i.e., $Q_k=X_k$ for $k \in A$, and
decodes no information sent over the reversely degraded
subchannels, i.e., $Q_k=\phi$ for $k \in A^c$. Hence for the
subchannels with $k \in A^c$, the link from the source node to the
relay node can be eliminated without changing the capacity of the
channel.

However, the relay node still plays an important role in the
reversely degraded subchannels by forwarding information that it
has decoded in other degraded subchannels to the destination node.
This is different from the role of the relay node in a single
reversely degraded channel, where it does not forward information
at all. Furthermore, we see that in the parallel relay channel,
information may be transmitted from the source node to the relay
node in one subchannel, and be forwarded to the destination node
over other subchannels, as we have commented in Remark
\ref{rk:crossflow}. More importantly, in contrast to the parallel
point-to-point channel, the capacity of the parallel relay channel
with degraded subchannels in Theorem \ref{th:dgcapa} can be larger
than the following sum of the capacities of the subchannels
\begin{equation}
\begin{split}
\max \; & \min \left\{ \sum_{k\in A}I(X_k,X_{Rk};Y_k),\;
\sum_{k\in
A}I(X_k;Y_{Rk}|X_{Rk})\right\} \\
& +\sum_{k\in A^c}I(X_k;Y_k|X_{Rk}).
\end{split}
\end{equation}
This demonstrates that the parallel relay channel is not a simple
combination of independent subchannels. This fact has also been
pointed out in \cite[Remark~15]{Liang0606it} for two-subchannel
case.

%\subsection{Gaussian Parallel Relay Channel with Degraded
%Subchannels}\label{sec:gausspara}

We now consider a Gaussian example of the parallel relay channel
with degraded subchannels. The channel input-output relationship
at one time instant is as follows.
\begin{equation}\label{eq:gausspara1}
\begin{split}
\text{For } k
\in A, \pp & Y_{Rk} = X_k+Z_{Rk} \\
& Y_k =X_k+ \sqrt{\rho_{Rk}}X_{Rk}+Z_{Rk}+Z'_k, \pp
\end{split}
\end{equation}
where $Z_{Rk}$ and $Z'_k$ are independent Gaussian random
variables with variances $\sigma_{Rk}^2$ and
$\sigma_k^2-\sigma_{Rk}^2$, respectively. For $k \in A$,
$\sigma_k^2> \sigma_{Rk}^2$.
\begin{equation}\label{eq:gausspara2}
\begin{split}
\text{For } k \in A^c, \pp & Y_{Rk} = X_k+Z_k+Z'_{Rk}  \\
& Y_k =X_k+ \sqrt{\rho_{Rk}}X_{Rk}+Z_k, \pp
\end{split}
\end{equation}
where $Z_k$ and $Z'_{Rk}$ are independent Gaussian random
variables with variances $\sigma_k^2$ and
$\sigma_{Rk}^2-\sigma_k^2$, respectively.  For $k \in A^c$,
$\sigma_{Rk}^2 \ge \sigma_k^2$. In \eqref{eq:gausspara1} and
\eqref{eq:gausspara2}, $\rho_{Rk}$ (assumed to be positive)
indicates the ratio of the relay-to-destination SNR to the
source-to-destination SNR for subchannel $k$. We assume that the
source and relay input sequences are subject to the following
average power constraints:
\begin{equation}
\frac{1}{n}\sum_{i=1}^n \sum_{k=1}^K \mE \left[X_{ki}^2\right]
\leq P \; , \hspace{4mm} \text{and} \hspace{4mm}
\frac{1}{n}\sum_{i=1}^n \sum_{k=1}^K \mE \left[ X_{Rki}^2\right]
\leq P_R \;.
\end{equation}
where $i$ is the time index.

It can be seen from \eqref{eq:gausspara1} and
\eqref{eq:gausspara2} that the subchannels with $k \in A$ satisfy
the degradedness condition \eqref{eq:dg1} and the subchannels with
$k \in A^c$ satisfy the degradedness condition \eqref{eq:dg2}.
Hence the Gaussian channel defined in \eqref{eq:gausspara1} and
\eqref{eq:gausspara2} is the parallel relay channel with degraded
subchannels. The following capacity theorem is based on Theorem
\ref{th:dgcapa}.
\begin{theorem}\label{th:gaucapa1}
The capacity of the Gaussian parallel relay channel with degraded
subchannels is given by
\begin{equation}\label{eq:gaucapa1}
\begin{split}
C=& \max_{\begin{array}{l} \scriptstyle \sum_{k=1}^K P_k \leq P,
\;\; \sum_{k=1}^K P_{Rk}\leq P_R, \\ \scriptstyle 0 \leq \beta_k
\leq 1,\; \text{ for }k=1,\ldots,K
\end{array}} \\
& \min \Bigg\{ \sum_{k=1}^K \cC
\left(\frac{P_k+\rho_{Rk}P_{Rk}+2\sqrt{\barbeta_k\rho_{Rk}P_kP_{Rk}}}{\sigma_k^2}\right),
\\
& \hspace{1cm} \sum_{k\in
A}\cC\left(\frac{\beta_kP_k}{\sigma_{Rk}^2}\right) +\sum_{k\in
A^c}\cC\left(\frac{\beta_kP_k}{\sigma_k^2}\right) \Bigg\}.
\end{split}
\end{equation}
where $\barbeta_k=1-\beta_k$, and the function
$\cC(x):=\frac{1}{2}\log (1+x)$.
\end{theorem}
In \eqref{eq:gaucapa1}, the parameter $\barbeta_k$ indicates
correlation between the source input and the relay input to
subchannel $k$, and $P_k$ and $P_{Rk}$ indicate the source and
relay powers that are allocated for transmission over subchannel
$k$.

\begin{proof}
The achievability follows from Theorem \ref{th:dgcapa} by choosing
the following joint distribution:
\begin{equation}
\begin{split}
& X_{Rk} \sim \cN(0, P_{Rk}), \\
& X'_k \sim \cN(0,\beta_k P_k), \pp \text{with $X'_k$ independent
of $X_{Rk}$}, \\
& X_k= \sqrt{\frac{\barbeta_k P_k}{P_{Rk}}}X_{Rk}+X'_k
\end{split}
\end{equation}
The converse is similar to the steps in the converse proof in
\cite[Sec.~IV]{Cover79}, and is omitted.
\end{proof}

Note that the capacity in Theorem \ref{th:gaucapa1} is sometimes
referred to as the synchronized capacity, because the source and
relay nodes are allowed to use correlated inputs to exploit
coherent combining gain. This may not be practical for encoder
design. It is hence interesting to study the asynchronized
capacity, where the source and relay nodes are assumed to use
independent inputs. The following asynchronized capacity is
derived by setting $\beta_k=1$ for $k=1,\ldots,K$ in
\eqref{eq:gaucapa1}.
\begin{corollary}\label{cor:gaucapa2}
For the Gaussian parallel relay channel with degraded subchannels,
the asynchronized capacity is given by
\begin{equation}\label{eq:gaucapa2}
\begin{split}
& C=\max_{\begin{array}{l} \scriptstyle \sum_{k=1}^K P_k \leq P, \\
\scriptstyle \sum_{k=1}^K P_{Rk}\leq P_R \end{array}} \min \Bigg\{
\sum_{k=1}^K \cC
\left(\frac{P_k+\rho_{Rk}P_{Rk}}{\sigma_k^2}\right), \\
& \hspace{2.8cm} \sum_{k\in
A}\cC\left(\frac{P_k}{\sigma_{Rk}^2}\right) +\sum_{k \in
A^c}\cC\left(\frac{P_k}{\sigma_k^2}\right) \Bigg\}.
\end{split}
\end{equation}
\end{corollary}

To obtain the capacity in Theorem \ref{th:gaucapa1} and the
asynchronized capacity in Corollary \ref{cor:gaucapa2}, we still
need to solve the optimization problems in \eqref{eq:gaucapa1} and
\eqref{eq:gaucapa2}, i.e., to find the jointly optimal correlation
parameters $\{\beta_k, \text{ for } k=1,\ldots,K\}$ and power
allocations $\{(P_k,P_{Rk}), \text{ for } k=1,\ldots,K \}$ in
\eqref{eq:gaucapa1}, and to find the optimal power allocations
$\{(P_k,P_{Rk}), \text{ for } k=1,\ldots,K \}$ in
\eqref{eq:gaucapa2}. We study these optimization problems in the
next section.
\section{Optimal Resource Allocation for Gaussian Parallel Relay
Channels with Degraded Subchannels}\label{sec:gauss}

In this section, we study the optimization problems in
\eqref{eq:gaucapa1} and \eqref{eq:gaucapa2}, which are {\it
max-min} optimization problems. We first introduce a general
technique for solving this class of {\it max-min} optimization
problems. We then demonstrate the application of this technique by
finding the optimal solutions in \eqref{eq:gaucapa1} and
\eqref{eq:gaucapa2}. We obtain the analytic form of the jointly
optimal correlation parameters $\{\beta_k, \text{ for }
k=1,\ldots,K\}$ and power allocation $\{(P_k,P_{Rk}), \text{ for }
k=1,\ldots,K \}$ that achieve the synchronized capacity for the
cases where the optimization problem is convex. We also obtain a
closed-form solution for the optimal $\{(P_k,P_{Rk}), \text{ for }
k=1,\ldots,K \}$ that achieve the asynchronized capacity. This
optimal solution may have three different structures depending on
the channel SNRs and power constraints. This optimal power
allocation is directly related to the power allocation for the
fading full-duplex relay channel presented in Section
\ref{sec:fullrelay}.

\subsection{Technique to Solve a Class of Max-Min Problem}

Consider the following max-min problem:
\begin{equation}\label{eq:maxmin}
\max_{\ut \in \cG} \; \min \left \{ R_1(\ut),\; R_2(\ut) \right \}
\end{equation}
where $\ut$ is a real vector in a set $\cG$, and $R_1(\ut)$ and
$R_2(\ut)$ are real continuous functions of $\ut$. An optimal
$\ut^*$ is referred to as a {\em max-min rule}.

We now introduce a technique to solve the max-min problem
\eqref{eq:maxmin}. We will also illustrate this technique with a
geometric interpretation. This technique is similar to that used
in finding the {\it minimax} detection rule in the two hypothesis
testing problem (see, e.g., \cite[Sec.~II.C]{Poor94}).
\begin{figure}[tbhp]
\begin{center}
\begin{psfrags}
\psfrag{v}[c]{$V(\alpha)$} \psfrag{a}[c]{$\textstyle
R\left(\alpha, \ut^{(\alpha_0)}\right)$} \psfrag{b}[c]{$R
\left(\alpha, \ut \right)$}
\psfrag{c}[c]{$R_2\left(\ut^{(\alpha_0)}\right)$}
\psfrag{e}[c]{$R_1\left(\ut^{(\alpha_0)}\right)$}
\psfrag{d}[c]{$R_2(\ut)$} \psfrag{f}[c]{$R_1(\ut)$}
\psfrag{h}[c]{$\alpha$} \psfrag{g}[c]{$\alpha_0$}
\epsfig{file=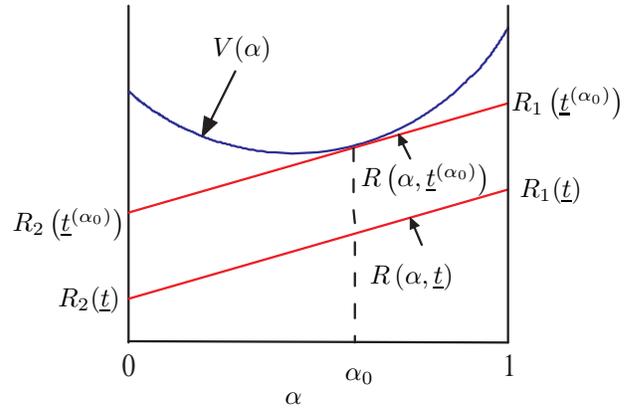,width=7cm}
\end{psfrags}
\caption{Illustration of functions $V(\alpha)$ and
$R(\alpha,\ut)$} \label{fig:vgraph}
\end{center}
\end{figure}

Consider the following function:
\begin{equation}
R(\alpha, \ut):=\alpha R_1(\ut)+(1-\alpha)R_2(\ut),\pp 0\leq
\alpha \leq 1.
\end{equation}
As a function of $\alpha$, $R(\alpha,\ut)$ is a straight line from
$R(0,\ut)=R_2(\ut)$ to $R(1,\ut)=R_1(\ut)$. Hence the maximization
in \eqref{eq:maxmin} corresponds to maximizing the minimal of the
two end points of the line $R(\alpha, \ut)$ over all possible $\ut
\in \cG$.

We further define a function
\begin{equation}
V(\alpha):=\max_{\ut \in \cG} R(\alpha,\ut)=R(\alpha,
\ut^{(\alpha)}),
\end{equation}
where $\ut^{(\alpha)}$ maximizes $R(\alpha,\ut)$ for fixed
$\alpha$. From the definitions of $V(\alpha)$ and $R(\alpha,\ut)$,
it is easy to see the following two facts (see
Fig.~\ref{fig:vgraph} for an illustration):
\begin{quote}
\renewcommand{\labelenumi}{Fact \theenumi:}
\begin{enumerate}
\item The function $V(\alpha)$ is continuous and convex for
$\alpha \in [0,1]$;

\item For any power allocation rule $\ut \in \cG$, $R(\alpha,\ut)$
as a function of $\alpha$ is completely below the convex curve
$V(\alpha)$ or tangent to it.
\end{enumerate}
\end{quote}

A known general solution to the max-min optimization problem in
\eqref{eq:maxmin} is summarized in the following proposition.

\begin{figure}[tbhp]
\begin{center}
\begin{tabular}{c}
\begin{psfrags}
\psfrag{v}[c]{$V(\alpha)$} \psfrag{a}[c]{$R\left(\alpha,
\ut^{(\alpha^*)}\right)$} \psfrag{c}[c]{$R_2(\ut^{(\alpha^*)})$}
\psfrag{e}[c]{$R_1(\ut^{(\alpha^*)})$}
\psfrag{h}[c]{$\alpha$}\psfrag{g}[c]{$\alpha^*=0$}
\epsfig{file=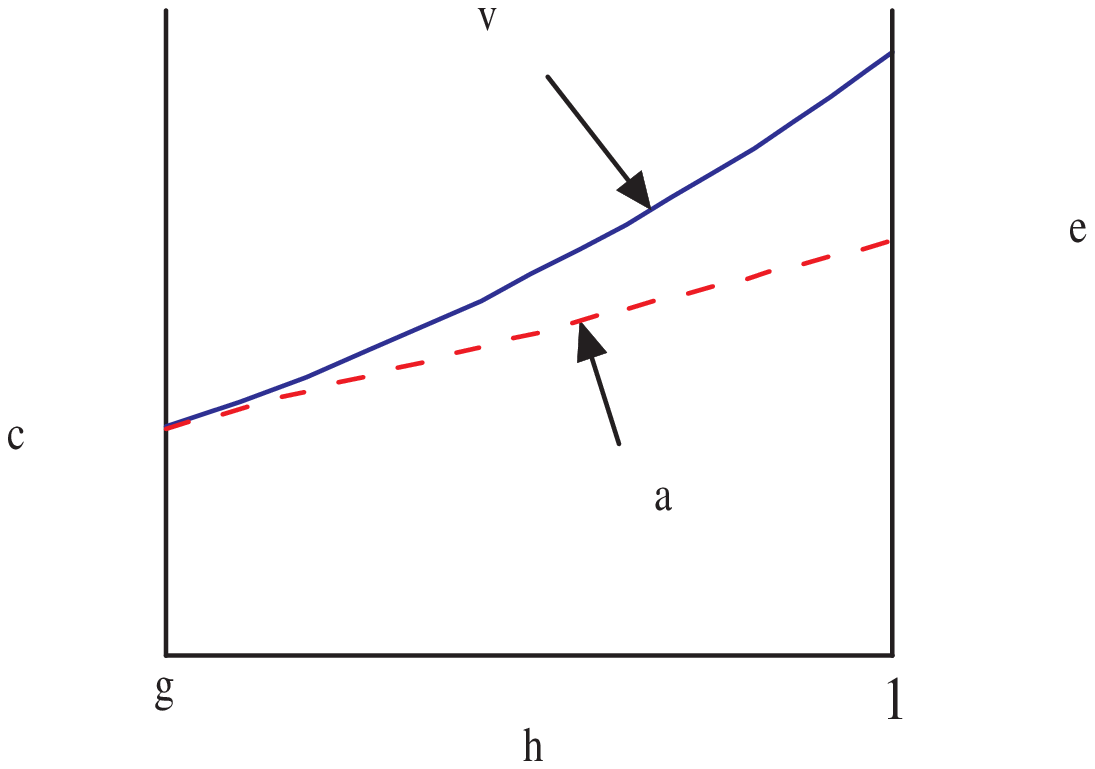,width=7cm}
\end{psfrags} \\
Case 1 \\ \\
\begin{psfrags}
\psfrag{v}[c]{$V(\alpha)$}
\psfrag{a}[c]{$R(\alpha,\ut^{(\alpha^*)})$}
\psfrag{c}[c]{$R_2(\ut^{(\alpha^*)})$}
\psfrag{e}[c]{$R_1(\ut^{(\alpha^*)})$} \psfrag{h}[c]{$\alpha$}
\psfrag{g}[c]{$\alpha^*=1$} \epsfig{file=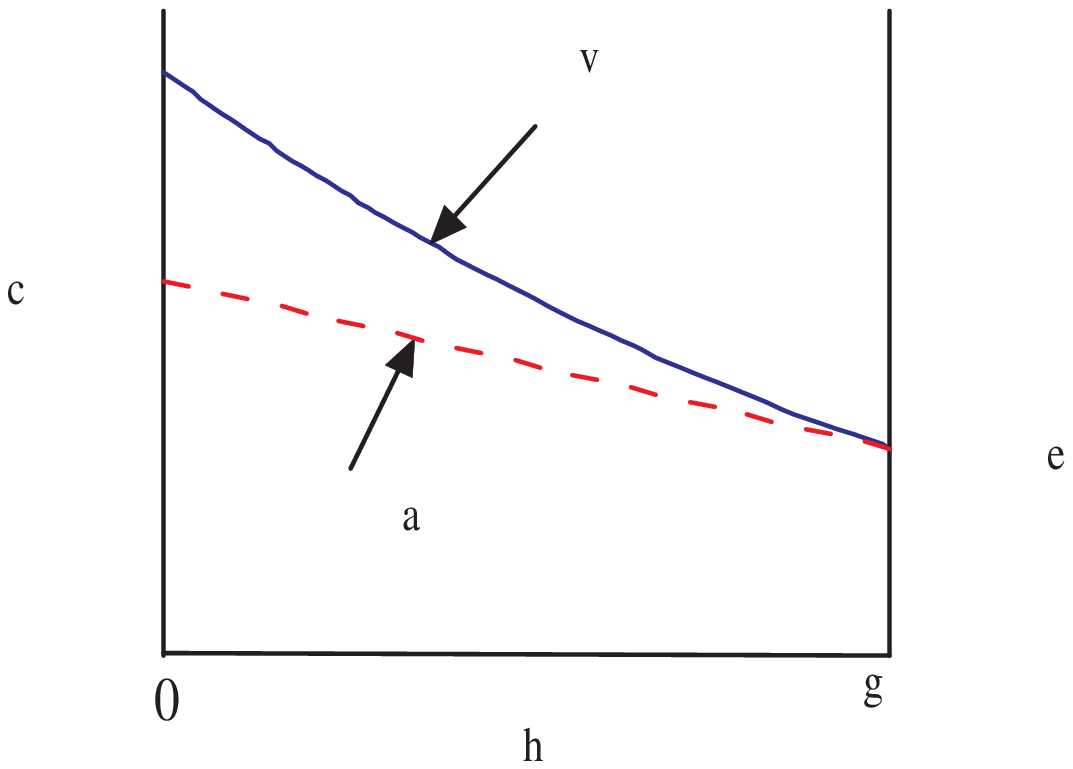,width=7cm}
\end{psfrags} \\
Case 2 \\ \\
\begin{psfrags} \psfrag{v}[c]{$V(\alpha)$} \psfrag{a}[c]{$ R(\alpha, \ut^{(\alpha^*)})$}
\psfrag{c}[c]{$R_2(\ut^{(\alpha^*)})$} \psfrag{e}[c]{
 $R_1(\ut^{(\alpha^*)})$} \psfrag{h}[c]{$\alpha$}
\psfrag{g}[c]{$\alpha^*$} \epsfig{file=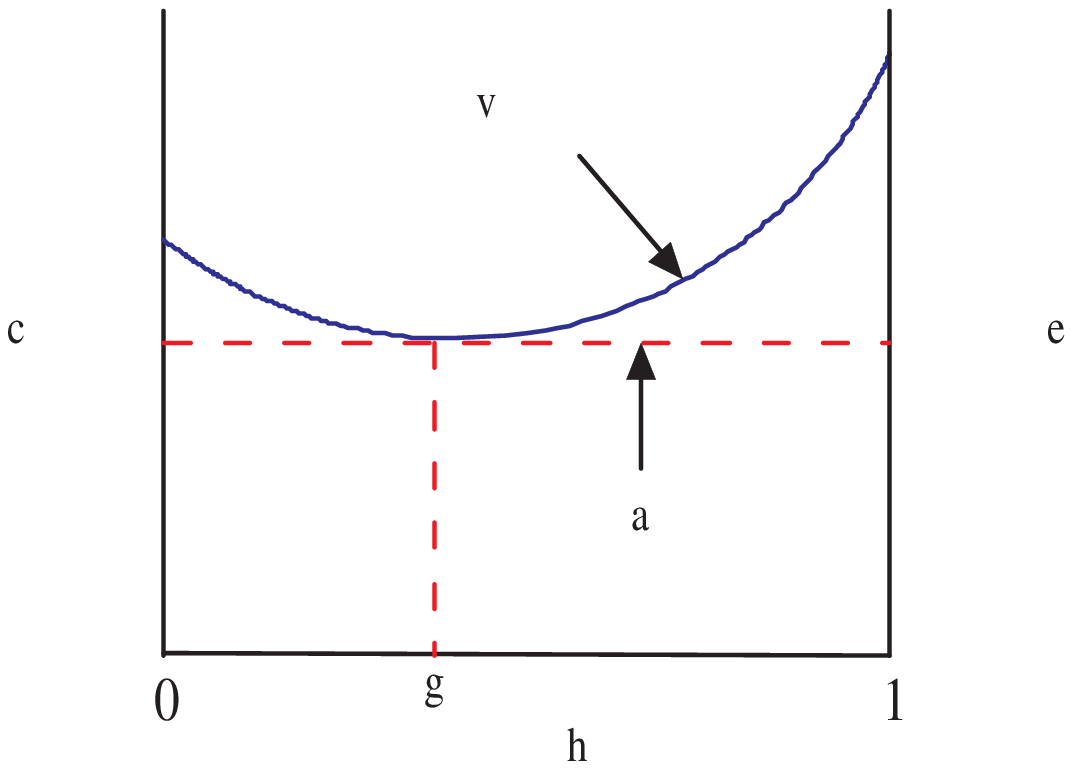,width=7cm}
\end{psfrags} \\
Case 3
\end{tabular}
\caption{Illustration of Cases 1, 2, and 3 in Proposition
\ref{prop:generule}} \label{fig:vcase}
\end{center}
\end{figure}

\begin{proposition}\label{prop:generule}
Suppose $\alpha^*$ is a solution to $V(\alpha^*)=\min_{\alpha \in
[0,1]} V(\alpha)$. Then $\ut^{(\alpha^*)}$ is a max-min rule,
i.e., a solution to the max-min problem in \eqref{eq:maxmin}. The
relationship between $R_1(\ut^{(\alpha^*)})$ and $R_2
(\ut^{(\alpha^*)})$ falls into the following three cases (see
Fig.~\ref{fig:vcase}):
\begin{quote}
\renewcommand{\labelenumi}{Case \theenumi:}
\begin{enumerate}
\item If $\alpha^*=0, \; R_1(\ut^{(\alpha^*)}) \ge R_2
(\ut^{(\alpha^*)})$;

\item If $\alpha^*=1,\; R_1(\ut^{(\alpha^*)}) \leq R_2
(\ut^{(\alpha^*)})$;

\item (Equalizer rule) If $0 < \alpha^* <1, \;
R_1(\ut^{(\alpha^*)}) = R_2 (\ut^{(\alpha^*)})$.
\end{enumerate}
\end{quote}
\end{proposition}

This technique of finding the max-min solution is applied
throughout the paper.
%The three cases in Proposition \ref{prop:generule} are illustrated
%in Figure \ref{fig:vcase}.
\subsection{Optimal Resource Allocation for Gaussian Parallel Relay Channel: Synchronized Case}

In this subsection, we apply Proposition \ref{prop:generule} to
find jointly optimal $\{ \beta_k, \text{ for } k=1,\ldots,K \}$
and $\{(P_k,P_{Rk}), \text{ for } k=1,\ldots,K \}$ that solve the
max-min problem in \eqref{eq:gaucapa1}. This optimal solution
provides the optimal correlation between the source and relay
inputs over each subchannel and the optimal source and relay power
allocation among the $K$ subchannels that achieve the synchronized
capacity of the Gaussian parallel relay channel with degraded
subchannels. We study the asynchronized case in the next
subsection.

To simplify notation, we let
\begin{equation}
\begin{split}
& \uP=(P_1,\ldots,P_K), \pp \uP_R=(P_{R1},\ldots,P_{RK}), \\
& \ubeta=(\beta_1,\ldots,\beta_K)
\end{split}
\end{equation}
and
\begin{equation}
\begin{split}
%& \ux=(\uP,\uP_R,\ubeta) \\
& \cG= \Bigg\{(\uP,\uP_R,\ubeta): \;\; \sum_{k=1}^K P_k \leq P,
\;\; \sum_{k=1}^K P_{Rk}\leq P_R, \\
& \hspace{1cm} 0 \leq \beta_k \leq 1, \;\; \text{for }
k=1,\ldots,K \Bigg\}
\end{split}
\end{equation}

The max-min optimization problem in \eqref{eq:gaucapa1} can be
written in the following compact form.
\begin{equation}\label{eq:gauopt1}
\fbox{$
\begin{split}
& C=\max_{(\uP,\uP_R,\ubeta) \in \cG } \min \{ \cR_1(\uP,\uP_R,\ubeta), \: \cR_2(\uP,\uP_R,\ubeta) \} \\
& \text{where}  \\
& \cR_1(\uP,\uP_R,\ubeta)\\
& \hspace{0.8cm} =\sum_{k=1}^K \cC
\left(\frac{P_k+\rho_{Rk}P_{Rk}+2\sqrt{\barbeta_k\rho_{Rk}P_kP_{Rk}}}{\sigma_k^2}\right)
\;
\\
& \cR_2(\uP,\uP_R,\ubeta)=\sum_{k\in
A}\cC\left(\frac{\beta_kP_k}{\sigma_{Rk}^2}\right) +\sum_{k\in
A^c}\cC\left(\frac{\beta_kP_k}{\sigma_k^2}\right)
\end{split}
$ }
\end{equation}

According to Proposition \ref{prop:generule}, the max-min rule
that solves \eqref{eq:gauopt1} may fall into the following three
cases.

\textbf{Case 1:} $\alpha^*=0$, and
$(\uP^{(0)},\uP_R^{(0)},\ubeta^{(0)})$ is a max-min rule, which
needs to satisfy the condition
\begin{equation}\label{eq:gaucon1}
R_1(\uP^{(0)},\uP_R^{(0)},\ubeta^{(0)}) \ge R_2
(\uP^{(0)},\uP_R^{(0)},\ubeta^{(0)}).
\end{equation}

By definition, $(\uP^{(0)},\uP_R^{(0)},\ubeta^{(0)})$ maximizes
\begin{equation}
R(0,\uP,\uP_R,\ubeta)=R_2(\uP,\uP_R,\ubeta).
\end{equation}
It is readily seen that the following $\ubeta^{(0)}$ is optimal:
\begin{equation}\label{eq:gaub1}
\beta^{(0)}_k=
\begin{cases}
1, \pp & \text{if } P^{(0)}_k >0; \\
\text{arbitrary} \pp & \text{if } P^{(0)}_k=0.
\end{cases}
\end{equation}
With $\ubeta^{(0)}$ given in \eqref{eq:gaub1},
$R_2(\uP,\uP_R,\ubeta)$ is a function of $\uP$ only. Moreover, it
is a convex function of $\uP$. Then the Kuhn-Tucker condition (KKT
condition) (see, e.g., \cite[p.~314-315]{Luen03}) characterizes
the necessary and sufficient condition that the optimal
$\uP^{(0)}$ needs to satisfy. The Lagrangian is given by
\begin{equation}
\cL = \sum_{k\in A}\cC\left(\frac{P_k}{\sigma_{Rk}^2}\right)
+\sum_{k\in A^c}\cC\left(\frac{P_k}{\sigma_k^2}\right) -\lambda
\left(\sum_{k=1}^K P_k-P \right),
\end{equation}
which implies the following KKT condition:
\begin{equation}
\begin{split}
& \frac{\partial \cL}{\partial P_k}=\frac{1}{2\ln
2}\cdot\frac{1}{\sigma_{Rk}^2+P_k} - \lambda \leq 0, \\
& \hspace{2.3cm} \text{with
equality if } P_k > 0, \spp \text{if } k \in A; \\
& \frac{\partial \cL}{\partial P_k}=\frac{1}{2\ln
2}\cdot\frac{1}{\sigma_k^2+P_k} - \lambda \leq 0, \\
& \hspace{2.3cm} \text{with equality if } P_k > 0, \spp \text{if }
k \in A^c.
\end{split}
\end{equation}
Hence the optimal $P_k^{(0)}$ is given by
\begin{equation}\label{eq:gaup1}
P_k^{(0)}=
\begin{cases}
\displaystyle \left( \frac{1}{2\ln 2\lambda}-\sigma_{Rk}^2 \right)^+ , \spp & \text{if } k \in A; \\
\displaystyle \left( \frac{1}{2\ln 2\lambda}-\sigma_k^2 \right)^+
, \spp & \text{if } k \in A^c
\end{cases}
\end{equation}
where $\lambda$ is chosen to satisfy the power constraint
$\sum_{k=1}^K P_k \leq P$. The function $(\cdot)^+$ is defined as
\begin{equation}
(x)^+ =
\begin{cases}
x \pp \mbox{if}\; x \ge 0 \\
0 \pp \mbox{if}\; x < 0.
\end{cases}
\end{equation}

For case 1 to happen, $(\uP^{(0)},\uP_R^{(0)},\ubeta^{(0)})$ needs
to satisfy the condition \eqref{eq:gaucon1}. To characterize the
least power $P_R$ needed for case 1 to happen, $\uP_R^{(0)}$ needs
to maximize $R_1 (\uP^{(0)},\uP_R, \ubeta^{(0)})$ with
$\ubeta^{(0)}$ given in \eqref{eq:gaub1} and $\uP^{(0)}$ given in
\eqref{eq:gaup1}, respectively. The optimal $\uP_R^{(0)}$ can be
obtained by the KKT condition via the following Lagrangian:
\begin{equation}\label{eq:lag}
\cL = \sum_{k=1}^K \cC
\left(\frac{P^{(0)}_k+\rho_{Rk}P_{Rk}}{\sigma_k^2}\right) -\mu
\left(\sum_{k=1}^K P_{Rk}-P_R \right),
\end{equation}
The KKT condition is given by
\begin{equation}
\begin{split}
& \frac{\partial \cL}{\partial P_{Rk}}=\frac{1}{2\ln
2}\cdot\frac{\rho_{Rk}}{\sigma_k^2+P^{(0)}_k+\rho_{Rk}P_{Rk}} -
\mu \leq 0, \\
& \hspace{3.5cm} \text{with equality if } P_{Rk} > 0
\end{split}
\end{equation}
which implies
\begin{equation}\label{eq:gaupr1}
P_{Rk}^{(0)}=\left(\frac{1}{2\ln 2
\mu}-\frac{P^{(0)}_k}{\rho_{Rk}}-\frac{\sigma_k^2}{\rho_{Rk}}
\right)^+ , \spp \text{for } k=1,\ldots,K
\end{equation}
where $\mu$ is chosen to satisfy the power constraint
$\sum_{k=1}^K P_{Rk} \leq P_R$.

Note that \eqref{eq:gaupr1} also follows directly from the
standard water-filling solution if we further derive
\eqref{eq:lag} in the following form:
\begin{equation}
\begin{split}
\cL = & \sum_{k=1}^K \cC
\left(\frac{P^{(0)}_k}{\sigma_k^2}\right)+\sum_{k=1}^K \cC
\left(\frac{\rho_{Rk}P_{Rk}}{P^{(0)}_k+\sigma_k^2}\right) \\
& -\mu \left(\sum_{k=1}^K P_{Rk}-P_R \right).
\end{split}
\end{equation}

With $\uP^{(0)}$, $\uP_{R}^{(0)}$, and $\ubeta^{(0)}$ given in
\eqref{eq:gaup1}, \eqref{eq:gaupr1}, and \eqref{eq:gaub1},
respectively, condition \eqref{eq:gaucon1} becomes
\begin{equation}\label{eq:gaucomp1}
\begin{split}
& \sum_{k=1}^K \cC
\left(\frac{P^{(0)}_k+\rho_{Rk}P^{(0)}_{Rk}}{\sigma_k^2}\right) \\
& \hspace{2cm} \ge \sum_{k\in
A}\cC\left(\frac{P^{(0)}_k}{\sigma_{Rk}^2}\right) +\sum_{k\in
A^c}\cC\left(\frac{P^{(0)}_k}{\sigma_k^2}\right)
\end{split}
\end{equation}
This condition is equivalent to the threshold condition $P_R \ge
P_{R,u}(P)$. The threshold $P_{R,u}(P)$ is a function of the
source power constraint $P$, and is determined by the value of
$P_R$ that results in equality in \eqref{eq:gaucomp1}.

Therefore, if case 1 occurs, the optimal source power allocation
$\uP^{(0)}$ has a {\em water-filling} form, and the optimal relay
power allocation $\uP_R^{(0)}$ also has a {\em water-filling} form
with $P^{(0)}_k+\sigma_k^2$ as the equivalent noise levels. The
optimal correlation parameter $\ubeta_k^{(0)}=1$ for
$P_k^{(0)}>0$, which indicates that coherent combining is not
needed for this case.

\textbf{Case 2:} $\alpha^*=1$, and
$(\uP^{(1)},\uP_R^{(1)},\ubeta^{(1)})$ is a max-min rule, which
needs to satisfy the condition
\begin{equation}\label{eq:gaucon2}
R_1(\uP^{(1)},\uP_R^{(1)},\ubeta^{(1)}) \leq R_2
(\uP^{(1)},\uP_R^{(1)},\ubeta^{(1)}).
\end{equation}

By definition, $(\uP^{(1)},\uP_R^{(1)},\ubeta^{(1)})$ maximizes
\begin{equation}
R(1,\uP,\uP_R,\ubeta)=R_1(\uP,\uP_R,\ubeta).
\end{equation}
We note that
\begin{equation}\label{eq:gaub2}
\beta^{(1)}_k=
\begin{cases}
0, \pp & \text{if } P^{(1)}_k > 0, \text{ and } P^{(1)}_{Rk}
> 0; \\
\text{arbitrary}, \pp & \text{otherwise}.
\end{cases}
\end{equation}

It can be shown that $R_1(\uP,\uP_R,\ubeta)$ is a convex function
of $(\uP,\uP_R)$ for $\ubeta^{(1)}$ given in \eqref{eq:gaub2}. To
derive the optimal $(\uP^{(1)},\uP^{(1)}_R)$ that maximizes
$R_1(\uP,\uP_R,\ubeta^{(1)})$, the Lagrangian can be written as
\begin{equation}
\begin{split}
\cL = & \sum_{k=1}^K \cC
\left(\frac{P_k+\rho_{Rk}P_{Rk}+2\sqrt{\rho_{Rk}P_kP_{Rk}}}{\sigma_k^2}\right)
\\
& -\lambda \left(\sum_{k=1}^K P_k-P \right)-\mu \left(\sum_{k=1}^K
P_{Rk}-P_R \right)
\end{split}
\end{equation}
The optimal $(\uP^{(1)},\uP_R^{(1)})$ needs to satisfy the
following KKT condition:
\begin{equation}\label{eq:gaukkt2}
\begin{split}
& \frac{\partial \cL}{\partial P_k}=\frac{1}{2\ln
2}\cdot\frac{\sqrt{P_k}+\sqrt{\rho_{Rk}P_{Rk}}}{\sigma_k^2+(\sqrt{P_k}+\sqrt{\rho_{Rk}P_{Rk}})^2}
\leq \lambda \sqrt{P_k}, \\
& \hspace{4.5cm} \text{with
equality if } P_k > 0; \\
& \frac{\partial \cL}{\partial P_{Rk}}=\frac{1}{2\ln
2}\cdot\frac{\sqrt{P_k}+\sqrt{\rho_{Rk}P_{Rk}}}{\sigma_k^2+(\sqrt{P_k}+\sqrt{\rho_{Rk}P_{Rk}})^2}
\leq \mu \sqrt{\frac{P_{Rk}}{\rho_{Rk}}}, \\
& \hspace{4.5cm} \text{with equality if } P_{Rk}
> 0.
\end{split}
\end{equation}
%Hence
%\begin{equation}\label{eq:gauppr2}
%\begin{split}
%P_k^{(1)}= \left( \frac{1}{\lambda (1+\frac{\lambda}{\mu})2\ln 2}
%-\frac{\sigma_{k}^2}{(1+\frac{\lambda}{\mu})^2}\right)^+ , \\
%P_{Rk}^{(1)}= \left( \frac{1}{\mu (1+\frac{\mu}{\lambda})2\ln 2}
%-\frac{\sigma_{k}^2}{(1+\frac{\mu}{\lambda})^2}\right)^+ ,
%\end{split}
%\end{equation}
%where $\lambda$ and $\mu$ are chosen to satisfy the power
%constraints $\sum_{k=1}^K P_k \leq P$ and $\sum_{k=1}^K P_{Rk}
%\leq P_R$.

From \eqref{eq:gaukkt2}, it is clear that $P^{(1)}_k=0 \iff
P^{(1)}_{Rk}=0$. According to \eqref{eq:gaub2}, we have
$\beta^{(1)}_k P^{(1)}_k=0$ for $k=1,\ldots,K$, which implies
$R_2(\uP^{(1)},\uP_R^{(1)},\ubeta^{(1)})=0$. Hence condition
\eqref{eq:gaucon2} cannot be satisfied. Therefore, case 2 never
happens.

\textbf{Case 3:} $0 < \alpha^* < 1$, and
$(\uP^{(\alpha^*)},\uP_R^{(\alpha^*)},\ubeta^{(\alpha^*)})$ is a
max-min rule, where $\alpha^*$ is determined by the following
condition
\begin{equation}\label{eq:gaucon3}
R_1(\uP^{(\alpha^*)},\uP_R^{(\alpha^*)},\ubeta^{(\alpha^*)}) =
R_2(\uP^{(\alpha^*)},\uP_R^{(\alpha^*)},\ubeta^{(\alpha^*)}).
\end{equation}
We need to derive
$(\uP^{(\alpha^*)},\uP_R^{(\alpha^*)},\ubeta^{(\alpha^*)})$ that
maximizes
\begin{equation}
R(\alpha^*,\uP,\uP_R,\ubeta)=\alpha^*
R_1(\uP,\uP_R,\ubeta)+(1-\alpha^*)R_2(\uP,\uP_R,\ubeta)
\end{equation}
for a fixed $\alpha^*$. This optimization problem is not convex.
Now the KKT condition provides only a necessary condition that the
optimal
$(\uP^{(\alpha^*)},\uP_R^{(\alpha^*)},\ubeta^{(\alpha^*)})$ needs
to satisfy. One can still perform a brute force search over those
$(\uP,\uP_R,\ubeta)$ that satisfy the KKT condition to find the
optimal
$(\uP^{(\alpha^*)},\uP_R^{(\alpha^*)},\ubeta^{(\alpha^*)})$.
However, it may be too complex to implement such an optimal
solution that involves designing correlated source and relay
inputs and also involves allocating the source and relay powers
jointly with the correlation parameter for each subchannel. Hence
it may not be worth searching for the jointly optimal solution
$(\uP^{(\alpha^*)},\uP_R^{(\alpha^*)},\ubeta^{(\alpha^*)})$,
except in case 1, where using independent source and relay inputs
is optimal and the optimal power allocation
$(\uP^{(\alpha^*)},\uP_R^{(\alpha^*)})$ is simpler. It is hence
more interesting to study the asynchronized case, where it is
assumed that the source and relay nodes use independent inputs.
\subsection{Optimal Resource Allocation for Gaussian Parallel Relay Channel: Asynchronized
Case}\label{sec:agauss}

In this subsection, we solve the max-min problem in
\eqref{eq:gaucapa2}. This problem is simpler than the max-min
problem in \eqref{eq:gaucapa1}, because the optimization is over
the power allocation $(\uP,\uP_R)$ only, and does not involve the
correlation parameters $\ubeta$. This also makes the optimal
solution easy to implement in practice. In the following, we fully
characterize the optimal power allocation, which may take three
possible structures.

We let
\begin{equation}
\cG= \left\{(\uP,\uP_R): \;\; \sum_{k=1}^K P_k \leq P, \;\;
\sum_{k=1}^K P_{Rk}\leq P_R \right\},
\end{equation}
and rewrite the max-min optimization problem in
\eqref{eq:gaucapa2} in the following manner:
\begin{equation}\label{eq:gauopt2}
\fbox{ $
\begin{split}
& C=\max_{(\uP,\uP_R)\in \cG } \min \{
\cR_1(\uP,\uP_R), \: \cR_2(\uP,\uP_R) \} \spp \\
& \text{where}\;\;   \\
& \cR_1(\uP,\uP_R)=\sum_{k=1}^K \cC
\left(\frac{P_k+\rho_{Rk}P_{Rk}}{\sigma_k^2}\right)
\\
& \cR_2(\uP,\uP_R)=\sum_{k\in
A}\cC\left(\frac{P_k}{\sigma_{Rk}^2}\right) +\sum_{k\in
A^c}\cC\left(\frac{P_k}{\sigma_k^2}\right)
\end{split}
$}
\end{equation}

We apply Proposition \ref{prop:generule} to solve
\eqref{eq:gauopt2}, and consider the following three cases.

\textbf{Case 1:} $\alpha^*=0$, and $(\uP^{(0)},\uP_R^{(0)})$ is a
max-min rule, which needs to satisfy the condition
\begin{equation}\label{eq:agaucon1}
R_1(\uP^{(0)},\uP_R^{(0)}) \ge R_2 (\uP^{(0)},\uP_R^{(0)}).
\end{equation}
The optimal $(\uP^{(0)},\uP_R^{(0)})$ can be derived following the
steps that are similar to those in case 1 of the synchronized
case, and is given by
\begin{equation}\label{eq:agauppr1}
\begin{split}
& P_k^{(0)}=
\begin{cases}
\displaystyle \left( \frac{1}{2\ln 2\lambda}-\sigma_{Rk}^2 \right)^+ , \spp & \text{if } k \in A \\
\displaystyle \left( \frac{1}{2\ln 2\lambda}-\sigma_k^2 \right)^+
, \spp & \text{if } k \in A^c
\end{cases} \\
& P_{Rk}^{(0)}=\left(\frac{1}{2\ln
2\mu}-\frac{P^{(0)}_k}{\rho_{Rk}}-\frac{\sigma_k^2}{\rho_{Rk}}
\right)^+ , \spp \text{for } k=1,\ldots,K
\end{split}
\end{equation}
where $\lambda$ and $\mu$ are chosen to satisfy the power
constraints $\sum_{k=1}^K P_k \leq P$ and $\sum_{k=1}^K P_{Rk}
\leq P_R$.

We refer to the optimal $(P_k^{(0)},P_{Rk}^{(0)})$ in
\eqref{eq:agauppr1} as {\em two-level water-filling} for the
following reason. The optimal $\uP^{(0)}$ is first obtained via
{\em water-filling} with respect to the noise levels
$\sigma_{Rk}^2$ and $\sigma_k^2$. The optimal $\uP^{(0)}_R$ is
then obtained via {\em water-filling} with
$P^{(0)}_k+\sigma_{k}^2$ as equivalent noise levels, where
$\uP^{(0)}$ is treated as an additional noise level.

With $(P_k^{(0)},P_{Rk}^{(0)})$ given in \eqref{eq:agauppr1},
condition \eqref{eq:agaucon1} becomes
\begin{equation}\label{eq:agaucomp1}
\begin{split}
& \sum_{k=1}^K \cC
\left(\frac{P^{(0)}_k+\rho_{Rk}P^{(0)}_{Rk}}{\sigma_k^2}\right) \\
& \hspace{2cm} \ge \sum_{k\in
A}\cC\left(\frac{P^{(0)}_k}{\sigma_{Rk}^2}\right) +\sum_{k\in
A^c}\cC\left(\frac{P^{(0)}_k}{\sigma_k^2}\right)
\end{split}
\end{equation}
This condition is equivalent to the threshold condition $P_R \ge
P_{R,u}(P)$, where the threshold $P_{R,u}(P)$ is determined by the
value of $P_R$ that results in equality in \eqref{eq:agaucomp1}.
The threshold $P_{R,u}(P)$ is clearly a function of the source
power constraint $P$.

\textbf{Case 2:} $\alpha^*=1$, and $(\uP^{(1)},\uP_R^{(1)})$ is a
max-min rule, which needs to satisfy the condition
\begin{equation}\label{eq:agaucon2}
R_1(\uP^{(1)},\uP_R^{(1)}) \leq R_2 (\uP^{(1)},\uP_R^{(1)}).
\end{equation}

By definition, $(\uP^{(1)},\uP_R^{(1)})$ maximizes
\begin{equation}
R(1,\uP,\uP_R)=R_1(\uP,\uP_R).
\end{equation}
We first note that $R_1(\uP,\uP_R)$ is a convex function of
$(\uP,\uP_R)$. The Lagrangian can be written as
\begin{equation}
\begin{split}
\cL = & \sum_{k=1}^K \cC
\left(\frac{P_k+\rho_{Rk}P_{Rk}}{\sigma_k^2}\right)-\lambda
\left(\sum_{k=1}^K P_k-P \right) \\
& -\mu \left(\sum_{k=1}^K P_{Rk}-P_R \right)
\end{split}
\end{equation}
According to the KKT condition, $(\uP^{(1)},\uP_R^{(1)})$ needs to
satisfy
\begin{equation}\label{eq:agaukkt2}
\begin{split}
& \frac{\partial \cL}{\partial P_k}=\frac{1}{2\ln
2}\cdot\frac{1}{\sigma_k^2+P_k+\rho_{Rk}P_{Rk}} \leq \lambda, \\
& \hspace{3cm} \text{with equality if } P_k > 0; \\
& \frac{\partial \cL}{\partial P_{Rk}}=\frac{1}{2\ln
2}\cdot\frac{1}{\sigma_k^2+P_k+\rho_{Rk}P_{Rk}} \leq
\frac{\mu}{\rho_{Rk}}, \\
& \hspace{3cm} \text{with equality if } P_{Rk} > 0
\end{split}
\end{equation}
which implies
\begin{equation}\label{eq:agauppr2}
\begin{split}
& \text{If } \lambda < \frac{\mu}{\rho_{Rk}}, \spp
P_k=\left(\frac{1}{2\ln 2\lambda}-\sigma_k^2 \right)^+,\spp P_{Rk}=0, \\
& \text{If } \lambda > \frac{\mu}{\rho_{Rk}}, \spp P_k=0, \spp
P_{Rk}=\left(\frac{1}{2\ln2 \mu}-\frac{\sigma_k^2}{\rho_{Rk}} \right)^+, \\
& \text{If } \lambda = \frac{\mu}{\rho_{Rk}}, \spp
P_k+\rho_{Rk}P_{Rk}=\left(\frac{1}{2\ln 2\lambda}-\sigma_k^2
\right)^+
\end{split}
\end{equation}
where $\lambda$ and $\mu$ are chosen to satisfy the power
constraints. In general, $\lambda \neq \frac{\mu}{\rho_{Rk}}$. The
equation \eqref{eq:agauppr2} implies an {\em orthogonal division
water-filling} power allocation, i.e., for each subchannel, either
the source node or the relay node allocates a positive amount of
power. This power allocation is similar to the optimal power
allocation for fading multiple access channels \cite{Knopp95}.

For case 2 to happen, $(\uP^{(1)},\uP_R^{(1)})$ needs to satisfy
the condition \eqref{eq:agaucon2}, i.e.,
\begin{equation}\label{eq:agaucomp2}
\begin{split}
& \sum_{k=1}^K \cC
\left(\frac{P^{(1)}_k+\rho_{Rk}P^{(1)}_{Rk}}{\sigma_k^2}\right) \\
& \hspace{2cm}\leq \sum_{k\in
A}\cC\left(\frac{P^{(1)}_k}{\sigma_{Rk}^2}\right) +\sum_{k\in
A^c}\cC\left(\frac{P^{(1)}_k}{\sigma_k^2}\right)
\end{split}
\end{equation}
This condition essentially requires that the relay power $P_R$ is
small compared to the source power $P$.

\textbf{Case 3:} $0 < \alpha^* < 1$, and
$(\uP^{(\alpha^*)},\uP_R^{(\alpha^*)})$ is a max-min rule, where
$\alpha^*$ is determined by the condition
\begin{equation}\label{eq:agaucon3}
R_1(\uP^{(\alpha^*)},\uP_R^{(\alpha^*)}) =
R_2(\uP^{(\alpha^*)},\uP_R^{(\alpha^*)}).
\end{equation}

We first derive $(\uP^{(\alpha^*)},\uP_R^{(\alpha^*)})$ that
maximizes
\begin{equation}\label{eq:ralpha}
R(\alpha^*,\uP,\uP_R)=\alpha^*
R_1(\uP,\uP_R)+(1-\alpha^*)R_2(\uP,\uP_R).
\end{equation}
for a given $\alpha^*$, and $\alpha^*$ will be determined later.

The Lagrangian can be written as
\begin{equation}
\begin{split}
\cL = & \alpha^* \sum_{k=1}^K \cC
\left(\frac{P_k+\rho_{Rk}P_{Rk}}{\sigma_k^2}\right)+ (1-\alpha^*)
\sum_{k\in A}\cC\left(\frac{P_k}{\sigma_{Rk}^2}\right) \\
& +(1-\alpha^*) \sum_{k\in
A^c}\cC\left(\frac{P_k}{\sigma_k^2}\right)-\lambda
\left(\sum_{k=1}^K P_k-P \right)\\
& -\mu \left(\sum_{k=1}^K P_{Rk}-P_R \right)
\end{split}
\end{equation}
which implies the following KKT condition:
\begin{equation}\label{eq:agaukkt31}
\begin{split}
\text{For } k \in A, \spp & \frac{\partial \cL}{\partial
P_k}=\frac{\alpha^*}{2\ln
2}\cdot\frac{1}{\sigma_k^2+P_k+\rho_{Rk}P_{Rk}}\\
& \hspace{1.2cm} + \frac{1-\alpha^*}{2\ln
2}\cdot\frac{1}{\sigma_{Rk}^2+P_k} \leq
\lambda ,\\
& \hspace{1.2cm} \text{ with equality if } P_k > 0;
\end{split}
\end{equation}
\begin{equation}\label{eq:agaukkt32}
\begin{split}
\text{For } k \in A^c, \spp & \frac{\partial \cL}{\partial
P_k}=\frac{\alpha^*}{2\ln
2}\cdot\frac{1}{\sigma_k^2+P_k+\rho_{Rk}P_{Rk}} \\
& \hspace{1.2cm} + \frac{1-\alpha^*}{2\ln
2}\cdot\frac{1}{\sigma_k^2+P_k} \leq
\lambda, \\
& \hspace{1.2cm} \text{ with equality if } P_k > 0;
\end{split}
\end{equation}
\begin{equation}\label{eq:agaukkt33}
\begin{split}
\text{For } k=1,\ldots,K, \; & \frac{\partial \cL}{\partial
P_{Rk}}=\frac{\alpha^*}{2\ln
2}\cdot\frac{\rho_{Rk}}{\sigma_k^2+P_k+\rho_{Rk}P_{Rk}} \leq \mu,
\\
& \hspace{1.2cm} \text{with equality if } P_{Rk} > 0.
\end{split}
\end{equation}
The optimal $(\uP^{(\alpha^*)},\uP_R^{(\alpha^*)})$ can be solved
by an iterative algorithm. For a given $\uP_R$, the value of $\uP$
can be obtained by solving \eqref{eq:agaukkt31} and
\eqref{eq:agaukkt32}, and its components have the following form:
\begin{equation}\label{eq:agaup3}
P_k=
\begin{cases}
\text{positive root $x$ of \eqref{eq:root31} if it exists,
otherwise } 0, \\
\hspace{5.4cm} \text{if}\;\; k \in A; \\
\text{positive root $x$ of \eqref{eq:root32} if it exists,
otherwise } 0, \\
\hspace{5.4cm} \text{if}\;\; k \in A^c \\
\end{cases}
\end{equation}
where the roots are determined by the following equations:
\begin{equation}\label{eq:root31}
\frac{\alpha^*}{2\ln
2}\cdot\frac{1}{x+\rho_{Rk}P_{Rk}+\sigma_k^2}+
\frac{1-\alpha^*}{2\ln 2}\cdot\frac{1}{x+\sigma_{Rk}^2} = \lambda
\end{equation}
\begin{equation}\label{eq:root32}
\frac{\alpha^*}{2\ln 2}\cdot\frac{1}{x+\rho_{Rk}P_{Rk}+
\sigma_k^2}+ \frac{1-\alpha^*}{2\ln 2}\cdot\frac{1}{x+\sigma_k^2}
= \lambda
\end{equation}
where $\lambda$ is chosen to satisfy the power constraint
$\sum_{k=1}^K P_k \leq P$. For a given $\uP$, the value of $\uP_R$
can be obtained by using \eqref{eq:agaukkt33}, and its components
have the following form:
\begin{equation}\label{eq:agaupr3}
P_{Rk}=\left(\frac{\alpha^*}{2\ln
2\mu}-\frac{P_k}{\rho_{Rk}}-\frac{\sigma_k^2}{\rho_{Rk}} \right)^+
, \spp \text{for } k=1,\ldots,K
\end{equation}
where $\mu$ is chosen to satisfy the power constraint
$\sum_{k=1}^K P_{Rk} \leq P_R$.

If we iteratively obtain $\uP$ and $\uP_R$ according to
\eqref{eq:agaup3} and \eqref{eq:agaupr3} with an initial $\uP_R$,
we show in the following that $(\uP,\uP_R)$ converges to an
optimal $(\uP^{(\alpha^*)},\uP^{(\alpha^*)}_R)$. We refer to this
optimal power allocation as the {\em iterative water-filling}
power allocation. We finally need to search over $0 \leq \alpha
\leq 1$ to find $\alpha^*$ that satisfies the equalizer condition
\eqref{eq:agaucon3}.

\begin{myproof1}
We show that $(\uP,\uP_R)$ obtained iteratively according to
\eqref{eq:agaup3} and \eqref{eq:agaupr3} converges to an optimal
$(\uP^{(\alpha^*)},\uP^{(\alpha^*)}_R)$. We first note that after
each iteration the objective function \eqref{eq:ralpha} either
increases or remains the same. We also note that the objective
function is bounded from the above because of the power
constraints at the source and relay nodes. Hence the objective
function must converge. It is easy to check that for a given
$\uP$, the objective function is a strictly concave function of
$\uP_R$, and \eqref{eq:agaupr3} yields the unique optimal $\uP_R$.
It is also true that for a fixed $\uP_R$, \eqref{eq:agaup3} yields
the unique optimal $\uP$. Hence as the objective function
converges, $(\uP,\uP_R)$ must converge. Moreover, $(\uP,\uP_R)$
converges to the solution of the KKT conditions, which are
sufficient for $(\uP,\uP_R)$ to be optimal because the objective
function is concave over $(\uP,\uP_R) \in \cG$.
\end{myproof1}

We now summarize the optimal power allocation that solves
\eqref{eq:gauopt2} in the following theorem.
\begin{theorem}
The optimal solution to \eqref{eq:gauopt2}, i.e., the optimal
power allocation that achieves the asynchronized capacity
\eqref{eq:gaucapa2} falls into the following three cases:

\textbf{Case 1}: The optimal $(\uP,\uP_R)$ takes the {\em
two-level water-filling} form and is given by \eqref{eq:agauppr1}.
This case happens if $P_R > P_{R,u}(P)$ where the threshold
$P_{R,u}(P)$ is determined by equality of \eqref{eq:agaucomp1}.

\textbf{Case 2}: The optimal $(\uP,\uP_R)$ takes the {\em
orthogonal division water-filling} form and is given by
\eqref{eq:agauppr2}. This case happens if condition
\eqref{eq:agaucomp2} is satisfied.

\textbf{Case 3}: The optimal $(\uP,\uP_R)$ takes the {\em
iterative water-filling} form and is obtained iteratively by
\eqref{eq:agaup3} and \eqref{eq:agaupr3}.
\end{theorem}

\section{Fading Full-Duplex Relay Channels}\label{sec:fullrelay}

\begin{figure}[tbhp]
\begin{center}
\includegraphics[width=8cm]{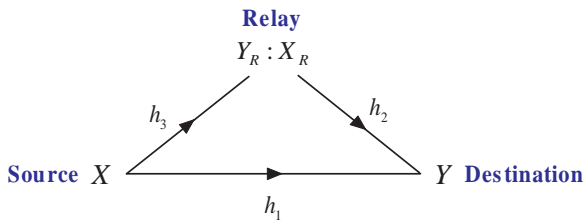}
\caption{Fading full-duplex relay channel} \label{fig:fullrelay}
\end{center}
\end{figure}

In this section, we study the three-terminal relay channel
\cite{Meul71,Cover79} in the context of wireless networks, where
nodes communicate over time-varying wireless channels. We are
interested in how the source and relay nodes should dynamically
change their power with wireless channel variation to achieve
optimal performance. Such wireless relay channels can be modelled
by the fading full-duplex relay model, where each transmission
link of a three-terminal relay channel \cite{Meul71,Cover79} is
corrupted by a multiplicative fading gain coefficient in addition
to an additive white Gaussian noise term (see
Fig.~\ref{fig:fullrelay}). The fading relay channel is referred to
as the {\em full-duplex} channel because the relay node is allowed
to transmit and receive at the same time and in the same frequency
band.

The channel input-output relationship at each symbol time can be
written as
\begin{equation}\label{eq:fullrelay}
\begin{split}
& Y=\sqrt{\rho_1}\: h_1 X + \sqrt{\rho_2}\: h_2 X_R+Z, \\
& Y_R=\sqrt{\rho_3}\: h_3 X+Z_R,
\end{split}
\end{equation}
where $h_1$, $h_2$, and $h_3$ are fading gain coefficients
corresponding to the three transmission links, respectively, and
are assumed to be independent complex proper random variables (not
necessarily Gaussian) with variances normalized to 1. We further
assume that the fading processes $\{h_{1i}\}$, $\{h_{2i}\}$, and
$\{h_{3i}\}$ are stationary and ergodic over time, where $i$ is
the time index. In \eqref{eq:fullrelay}, the additive noise terms
$Z$ and $Z_R$ are independent proper complex Gaussian random
variables with variances also normalized to 1. The parameters
$\rho_1$, $\rho_2$, and $\rho_3$ represent the link gain to noise
ratios of the corresponding transmission links. The input symbol
sequences $\{X_i\}$ and $\{ X_{Ri}\}$ are subject to separate
average power constraints $P$ and $P_R$, respectively, i.e.,
\begin{equation}\label{eq:powercon}
\frac{1}{n}\sum_{i=1}^n \mE|X_i|^2 \leq P, \hspace{1cm}
\frac{1}{n}\sum_{i=1}^n \mE|X_{Ri}|^2 \leq P_R.
\end{equation}

\begin{remark}
The fading relay channel is a special case of the parallel relay
channel with each subchannel corresponding to one fading state
realization. In particular, for a given fading state the fading
relay channel is a Gaussian relay channel by \eqref{eq:fullrelay}.
However, since this Gaussian channel is not physically degraded,
the fading relay channel is not a Gaussian parallel relay channel
with degraded subchannels that is considered in Sections
\ref{sec:pararelay} and \ref{sec:gauss}, where physically
degradedness is assumed for each subchannel.
\end{remark}

We assume that the transmitter and the receiver know the channel
state information instantly. Hence the source and relay nodes can
allocate their transmitted signal powers according to the channel
state information to achieve the best performance. Our goal is to
study the optimal power allocation at the source and relay nodes.
As in Section \ref{sec:agauss}, we are interested in the
asynchronized case for the fading full-duplex relay channel, where
the source and relay nodes are required to use independent inputs.
The main reason is because this simplifies the transmitter design,
and is more practical in distributed networks, where nodes need to
construct their codebooks independently.

%We further note that the fading full-duplex relay channel is
%similar to the Gaussian relay channel with degraded subchannels
%considered in the preceding section in that each subchannel
%corresponds to the fading channel under one given fading state.
%The difference between the two channels is that the fading channel
%under one fading state does not satisfy the degradedness condition
%\eqref{eq:dgcon}.

%The capacity of the fading full-duplex relay channel is still an
%open problem. Lower and upper bounds on the capacity have been
%given in \cite{Host05}, where the resource allocations that
%optimizes these bounds were derived under a sum power constraint,
%i.e., the source and relay nodes are subject to a total power
%constraint. In this paper, we study the power allocations that
%optimize the capacity bounds with the source node and relay node
%subject to separate power constraints as given in
%\eqref{eq:powercon}. In particular, we characterize the conditions
%where the lower and upper bounds match and determine the capacity
%of the channel.

For notational convenience, we collect the fading coefficients
$h_1$, $h_2$ and $h_3$ in a vector $\uh:=(h_1,h_2,h_3)$. We define
a set $A:= \{\uh: \rho_3|h_3|^2
> \rho_1 |h_1|^2 \}$, which contains all the fading states $\uh$
with the source-to-relay link being better than the
source-to-destination link. The complement of the set $A$ is
$A^c:= \{\uh: \rho_3 |h_3|^2 \leq \rho_1 |h_1|^2 \}$. We define a
set $\cG$ that contains all power allocation functions that
satisfy the power constraints, i.e.,
\begin{equation}\label{eq:fullcg}
\cG=\left\{ (P(\uh),P_R(\uh)): \mE[ P(\uh)] \leq P, \;\; \mE[
P_R(\uh)] \leq P_R \right\}.
\end{equation}

The following lower and upper bounds on the asynchronized capacity
of the fading full-duplex relay channel were given in
\cite{Host05}.
\begin{lemma}\label{lemma:fulllowup}(\cite{Host05})
For the fading full-duplex relay channel, lower and upper bounds
on the asynchronized capacity are given by
\begin{equation}\label{eq:fullalow}
\begin{split}
& C_{\mathtt{low}} =\max_{(P(\uh),P_R(\uh))\in \cG} \\
&  \min \Bigg \{
2\mE \left[ \cC\Big( P(\uh)\rho_1 |h_1|^2+ P_R(\uh)\rho_2|h_2|^2\Big)\right],\\
& \hspace{1cm} 2\mE_A
\left[\cC\Big(P(\uh)\rho_3|h_3|^2\Big)\right] +2\mE_{A^c}\left[
\cC\Big(P(\uh)\rho_1|h_1|^2\Big) \right] \Bigg \}
\end{split}
\end{equation}
\begin{equation}\label{eq:fullaup}
\begin{split}
C_{\mathtt{up}}= & \max_{(P(\uh),P_R(\uh))\in \cG} \\
& \hspace{0.7cm}\min \Bigg \{
2\mE \left[ \cC\Big( P(\uh)\rho_1 |h_1|^2+ P_R(\uh)\rho_2|h_2|^2\Big)\right],\\
& \hspace{1.7cm} 2\mE
\left[\cC\Big(P(\uh)(\rho_1|h_1|^2+\rho_3|h_3|^2)\Big)\right]
\Bigg \}
\end{split}
\end{equation}
\end{lemma}
Note that the rates in the lower bound of Lemma
\ref{lemma:fulllowup} are the same as the achievable rates in
Corollary \ref{cor:gaucapa2}.

The optimal power allocation that maximizes the lower bound
\eqref{eq:fullalow} and the upper bound \eqref{eq:fullaup} were
obtained in \cite{Host05} under a sum power constraint, i.e., the
source and relay nodes are subject to a total power constraint. In
this paper, we assume that the source and relay nodes are subject
to the separate power constraints as given in \eqref{eq:powercon}
and \eqref{eq:fullcg}, and derive the optimal power allocations
that maximize the bounds \eqref{eq:fullalow} and
\eqref{eq:fullaup}, respectively. We also characterize the
conditions where the lower and upper bounds match and determine
the capacity of the channel.

Using the same technique as in Section \ref{sec:agauss}, we
characterize the optimal power allocation that maximizes the lower
bound \eqref{eq:fullalow} of the fading relay channel. This
optimal power allocation takes the same three structures as those
given in Section \ref{sec:agauss}, and is summarized in the
following for the sake of completeness.

\begin{quote}
\textbf{Optimal power allocation that maximizes the lower bound
\eqref{eq:fullalow}:}

\vspace{-2mm}
 \rule{\linewidth}{0.3mm}

\begin{small}
\textbf{Case 1 (two-level water-filling):} If $P_R \ge
P_{R,u}(P)$, the optimal $(P^{(0)}(\uh),P_R^{(0)}(\uh))$ is given
by
\begin{equation}\label{eq:fullp1}
P^{(0)}(\uh)=
\begin{cases}
\displaystyle \left( \frac{1}{\lambda\ln 2}-\frac{1}{\rho_3
|h_3|^2} \right)^+ , &  \pp \text{if}\;\; \uh \in A,  \\
\displaystyle \left( \frac{1}{\lambda \ln 2}-\frac{1}{\rho_1
|h_1|^2} \right)^+ , &  \pp \text{if}\;\; \uh \in A^c
\end{cases}
\end{equation}
where $\lambda$ is chosen to satisfy the power constraint $\mE[
P(\uh)]=P$.

\begin{equation}\label{eq:fullpr1}
P_R^{(0)}(\uh)= \left( \frac{1}{\mu \ln 2}-\frac{1+\rho_1
|h_1|^2P^{(0)}(\uh)}{\rho_2|h_2|^2} \right)^+
\end{equation}
where $\mu$ is chosen to satisfy the power constraint $\mE[
P_R(\uh)]=P_R$.

The threshold $P_{R,u}(P)$ as a function of the source power $P$
can be solved using the following equation
\begin{equation}
\begin{split}
& \mE \left[\cC\Big( P^{(0)}(\uh)\rho_1 |h_1|^2+ P_R^{(0)}(\uh)\rho_2|h_2|^2\Big)\right]\\
& \hspace{1cm} =
\mE_A\left[\cC\Big(P^{(0)}(\uh)\rho_3|h_3|^2\Big)\right]
\\
& \hspace{1.4cm} +\mE_{A^c}\left[
\cC\Big(P^{(0)}(\uh)\rho_1|h_1|^2\Big) \right]
\end{split}
\end{equation}

\textbf{Case 2 (orthogonal division water-filling):} The optimal
$(P^{(1)}(\uh),P_R^{(1)}(\uh))$ is given by
\begin{equation}\label{eq:fullppr2}
\begin{split}
& \text{If }\; \frac{\lambda}{\rho_1 |h_1|^2}\leq
\frac{\mu}{\rho_2 |h_2|^2}, \\
& \spp P^{(1)}(\uh)= \left( \frac{1}{\lambda \ln
2}-\frac{1}{\rho_1 |h_1|^2} \right)^+, \spp P_R^{(1)}(\uh)=0; \\
& \text{If }\; \frac{\lambda}{\rho_1 |h_1|^2} > \frac{\mu}{\rho_2
|h_2|^2}, \\
& \spp P^{(1)}(\uh)=0, \spp P_R^{(1)}(\uh)=\left( \frac{1}{\mu \ln
2}-\frac{1}{\rho_2|h_2|^2} \right)^+
\end{split}
\end{equation}
where $\lambda$ and $\mu$ are chosen to satisfy the power
constraints $\mE[ P(\uh)]=P$ and $\mE[ P_R(\uh)]=P_R$.

Case 2 happens if the following condition is satisfied:
\begin{equation}\label{eq:fullcomp2}
\begin{split}
& \mE \left[\cC\Big( P^{(1)}(\uh)\rho_1 |h_1|^2+ P_R^{(1)}(\uh)\rho_2|h_2|^2\Big)\right],\\
& \pp \leq
\mE_A\left[\cC\Big(P^{(1)}(\uh)\rho_3|h_3|^2\Big)\right] \\
& \hspace{1.5cm}+\mE_{A^c}\left[
\cC\Big(P^{(1)}(\uh)\rho_1|h_1|^2\Big) \right]
\end{split}
\end{equation}

\textbf{Case 3 (iterative water-filling):} The optimal
$(P^{(\alpha^*)},P_R^{(\alpha^*)})$ can be obtained by the
following iterative algorithm. For a given $P_R(\uh)$, the value
of $P(\uh)$ is given by
\begin{equation}\label{eq:fullp3}
P(\uh)=
\begin{cases}
\text{positive root $x$ of \eqref{eq:rootfull1} if it exists,
otherwise } 0, \\
\hspace{4.5cm} \text{if}\;\; \uh \in A; \\
\text{positive root $x$ of \eqref{eq:rootfull2} if it exists,
otherwise } 0, \\
\hspace{4.5cm} \text{if}\;\; \uh \in A^c
\end{cases}
\end{equation}
where the roots are determined by the following equations:
\begin{equation}\label{eq:rootfull1}
\begin{split}
& \frac{\alpha^*}{\ln 2}\cdot\frac{\rho_1 |h_1|^2}{\rho_1
|h_1|^2x+ P_R(\uh)\rho_2|h_2|^2+1} \\
& \hspace{1.4cm} + \frac{1-\alpha^*}{\ln
2}\cdot\frac{\rho_3|h_3|^2}{1+\rho_3|h_3|^2x} = \lambda
\end{split}
\end{equation}
\begin{equation}\label{eq:rootfull2}
\begin{split}
& \frac{\alpha^*}{\ln 2}\cdot\frac{\rho_1 |h_1|^2}{\rho_1
|h_1|^2x+ P_R(\uh)\rho_2|h_2|^2+1} \\
& \hspace{1.4cm}+ \frac{1-\alpha^*}{\ln
2}\cdot\frac{\rho_1|h_1|^2}{1+\rho_1|h_1|^2x} = \lambda
\end{split}
\end{equation}
where $\lambda$ is chosen to satisfy the power constraint $\mE[
P(\uh)]=P$. For a given $P(\uh)$, the value of $P_R(\uh)$ is given
by
\begin{equation}\label{eq:fullpr3}
P_R(\uh)= \left( \frac{1}{\mu \ln 2}-\frac{1+\rho_1
|h_1|^2P(\uh)}{\rho_2|h_2|^2} \right)^+
\end{equation}
where $\mu$ is chosen to satisfy the power constraint $\mE[
P_R(\uh)]=P_R$.

The power allocation $(P(\uh),P_R(\uh))$ obtained iteratively from
\eqref{eq:fullp3} and \eqref{eq:fullpr3} with an initial
$P_R(\uh)$ converges to an optimal
$(P^{(\alpha^*)}(\uh),P^{(\alpha^*)}_R(\uh))$. The parameter
$\alpha^*$ is determined by the following equalizer condition:
\begin{equation}
\begin{split}
&\mE \left[ \cC\Big( P(\uh)\rho_1 |h_1|^2+
P_R(\uh)\rho_2|h_2|^2\Big)\right] \\
& \pp =\mE_A \left[\cC\Big(P(\uh)\rho_3|h_3|^2\Big)\right] \\
& \hspace{1.5cm} +\mE_{A^c}\left[ \cC\Big(P(\uh)\rho_1|h_1|^2\Big)
\right]
\end{split}
\end{equation}
\end{small}
\rule{\linewidth}{0.3mm}
\end{quote}

The optimal power allocation for the upper bound
\eqref{eq:fullaup} can be derived in a similar fashion but it is
omitted here since this optimization does not have an operation
meaning. In general, the upper and lower bounds do not match. In
the following theorem, we characterize the condition where the two
bounds match and establish the asynchronized capacity.
\begin{theorem}\label{th:fullacapa}
For the fading full-duplex relay channel, if the channel
statistics and the power constraints at the source and relay nodes
satisfy the condition \eqref{eq:fullcomp2}, then the asynchronized
capacity is given by
\begin{equation}\label{eq:fullacapa}
C =2\mE \left[ \cC\Big( P^{(1)}(\uh)\rho_1 |h_1|^2+ P^{(1)}_R(\uh)\rho_2|h_2|^2\Big)\right]\\
\end{equation}
where the capacity achieving power allocation
$(P^{(1)}(\uh),P_R^{(1)}(\uh))$ takes the orthogonal (time)
division water-filling form given in \eqref{eq:fullppr2}.
\end{theorem}
\begin{proof}
The lower bound \eqref{eq:fullalow} and the upper bound
\eqref{eq:fullaup} have one term in common inside the ``min" in
their expression. If condition \eqref{eq:fullcomp2} is satisfied,
case 2 happens when solving the max-min problem for the lower
bound \eqref{eq:fullalow}. In this case, the common term of the
bounds is optimized by the power allocation in \eqref{eq:fullppr2}
and determines both bounds that result in
$C_{\mathtt{up}}=C_{\mathtt{low}}$. This common value is thus the
asynchronized capacity.
\end{proof}

The condition given in Theorem \ref{th:fullacapa} essentially
requires that the relay power $P_R$ be small compared to the
source power $P$. In this case, the optimal scheme is to maximize
the rate at which the source and relay nodes can transmit to the
destination node. The optimal scheme is to let the source and
relay nodes have a time division access of the channel. For a
given channel state realization, the node with a better channel to
the destination node is allowed to transmit. This is similar to
the optimal power allocation scheme for the fading multiple access
channel studied in \cite{Knopp95}.
\section{Fading Half-Duplex Relay Channels}\label{sec:halfrelay}

\begin{figure}[tbhp]
\begin{center}
\includegraphics[width=8.5cm]{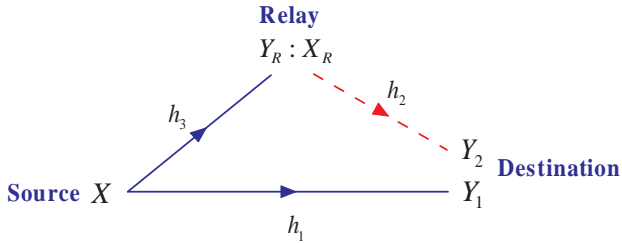}
\caption{Fading half-duplex relay model} \label{fig:halfrelay}
\end{center}
\end{figure}

In this section, we study a fading half-duplex relay channel
model, where the source node transmits to the relay and
destination nodes in one channel (channel 1), and the relay node
transmits to the destination node in an orthogonal channel
(channel 2). We introduce a parameter $\theta$ to represent the
channel resource (time and bandwidth) allocation between the two
orthogonal channels. We draw this fading half-duplex relay channel
model in Fig.~\ref{fig:halfrelay} with the solid and dashed lines
indicating the transmission links of channels 1 and 2,
respectively.

The input-output relationship for the fading half-duplex relay
channel is given by
\begin{equation}\label{eq:fdmodel}
\begin{split}
& Y_1=\sqrt{\rho_1}\: h_1 X+Z_1, \\
& Y_2=\sqrt{\rho_2}\: h_2 X_R+Z_2, \\
& Y_R=\sqrt{\rho_3}\: h_3 X+Z_R,
\end{split}
\end{equation}
where $h_1$, $h_2$, and $h_3$ are fading gain coefficients that
satisfy the same assumptions as for the fading full-duplex relay
channel in Section \ref{sec:fullrelay}. The additive noise terms
$Z_1$, $Z_2$, and $Z_R$ are independent proper complex Gaussian
random variables with variances normalized to 1. The parameters
$\rho_1$, $\rho_2$, and $\rho_3$ represent the link gain to noise
ratios of the corresponding transmission links. The source and
relay input sequences are subject to the same power constraints
\eqref{eq:powercon} as in the fading full-duplex relay channel.

As in the full-duplex case, the channel state information is
assumed to be known at both the transmitter and the receiver.
Hence the source and relay nodes can allocate their powers
adaptively according to the instantaneous channel state
information. The half-duplex channel has an additional channel
resource allocation parameter $\theta$ that may also be optimized.
Our goal is to find the jointly optimal $\theta$ and power
allocation for the source and relay nodes that achieve the best
rate. We also derive an upper bound on the capacity, which helps
to establish capacity theorems for some special cases.

We study three scenarios. In Scenario I, we fix
$\theta=\frac{1}{2}$, and only consider the maximization of the
achievable rate over the power allocation at the source and relay
nodes. In Scenario II, we restrict $\theta$ to be same for all
channel states, and jointly optimize the achievable rate over this
single parameter $\theta$ and power allocation. In Scenario III,
which is the most general scenario, we further allow $\theta$ to
change with channel state realizations, and optimize the
achievable rate over all possible channel resource and power
allocations.

\subsection{Scenario I: Fixed $\theta=1/2$}\label{sec:scenarioI}

In this subsection, we study Scenario I, where the two orthogonal
channels share the channel resource equally, i.e., the channel
resource allocation parameter $\theta=1/2$. We use this scenario
to demonstrate the three basic structures of the optimal power
allocation, which take simple forms. The optimal power allocation
can be implemented in a distributed manner at the source and relay
nodes, because each node needs to know only the channel state
information of the links over which it transmits.

In the following, we first give an achievable rate for this
channel, and then find an optimal power allocation that maximizes
this achievable rate.
\begin{proposition}\label{prop:fdlowI}
An achievable rate for the fading half-duplex relay channel
Scenario I is given by
\begin{equation}\label{eq:fdlowI}
\begin{split}
& C_{\mathtt{low}}= \max_{(P(\uh),P_R(\uh))\in \cG} \\
& \min \Bigg \{ \mE \left[ \cC\Big(2 P(\uh)\rho_1 |h_1|^2 \Big)+
\cC\Big(2 P_R(\uh)\rho_2|h_2|^2\Big)\right],\\
& \hspace{1cm} \mE_A \left[\cC\Big( 2
P(\uh)\rho_3|h_3|^2\Big)\right] +\mE_{A^c}\left[ \cC\Big(
2P(\uh)\rho_1|h_1|^2\Big) \right] \Bigg \}
\end{split}
\end{equation}
\end{proposition}

Proposition \ref{prop:fdlowI} follows easily by using steps that
are similar to the achievability proof for Theorems
\ref{th:dgcapa} and \ref{th:gaucapa1} and by using the channel
definition \eqref{eq:fdmodel}.

The optimal power allocation that maximizes $C_{\mathtt{low}}$ in
\eqref{eq:fdlowI} can be derived by applying Proposition
\ref{prop:generule}, and are given in the following three cases.
The details of the proof are relegated to Appendix
\ref{app:scenarioI}.

\begin{quote}
\textbf{Optimal power allocation that maximizes the lower bound
\eqref{eq:fdlowI}:}

\vspace{-2mm}
 \rule{\linewidth}{0.3mm}

\begin{small}
\textbf{Case 1:} If $P_R \ge P_{R,u}(P)$, the optimal
$(P^{(0)}(\uh),P_R^{(0)}(\uh))$ is given by
\begin{equation}\label{eq:fdp1}
P^{(0)}(\uh)=
\begin{cases}
\displaystyle \frac{1}{2}\left( \frac{1}{\lambda \ln
2}-\frac{1}{\rho_3 |h_3|^2} \right)^+ , &  \spp \text{if}\;\; \uh
\in A,  \\ \displaystyle \frac{1}{2}\left( \frac{1}{\lambda \ln
2}-\frac{1}{\rho_1 |h_1|^2} \right)^+ , &  \spp \text{if}\;\; \uh
\in A^c
\end{cases}
\end{equation}
where $\lambda$ is chosen to satisfy the power constraint $\mE[
P(\uh)]=P$.

\begin{equation}\label{eq:fdpr1}
P_R^{(0)}(\uh)= \frac{1}{2}\left( \frac{1}{\mu \ln
2}-\frac{1}{\rho_2|h_2|^2} \right)^+
\end{equation}
where $\mu$ is chosen to satisfy the power constraint $\mE[
P_R(\uh)]=P_R$.

The threshold $P_{R,u}(P)$ as a function of the source power $P$
can be solved using the following equation:
\begin{equation}\label{eq:compare1}
\begin{split}
& \mE \left[ \cC\Big(2 P_R^{(0)}(\uh)\rho_2|h_2|^2\Big)\right] \\
& =  \mE_A \left[ \cC\Big(2 P^{(0)}(\uh)\rho_3|h_3|^2\Big)-
\cC\Big(2 P^{(0)}(\uh)\rho_1|h_1|^2\Big) \right].
\end{split}
\end{equation}

\textbf{Case 2:} If $P_R \leq P_{R,l}(P)$, the optimal
$(P^{(1)}(\uh),P_R^{(1)}(\uh))$ is given by
\begin{equation}\label{eq:fdp2}
P^{(1)}(\uh)= \frac{1}{2}\left( \frac{1}{\lambda \ln
2}-\frac{1}{\rho_1 |h_1|^2} \right)^+
\end{equation}
\begin{equation}\label{eq:fdpr2}
P_R^{(1)}(\uh)= \frac{1}{2}\left( \frac{1}{\mu \ln
2}-\frac{1}{\rho_2|h_2|^2} \right)^+
\end{equation}
where $\lambda$ and $\mu$ are chosen to satisfy the power
constraints $\mE[ P(\uh)]=P$ and $\mE[ P_R(\uh)]=P_R$.

The threshold $P_{R,l}(P)$ can be solved using the following
equation:
\begin{equation}\label{eq:compare2}
\begin{split}
& \mE \left[ \cC\Big(2 P_R^{(1)}(\uh)\rho_2|h_2|^2\Big)\right] \\
& = \mE_A \left[ \cC\Big(2 P^{(1)}(\uh)\rho_3|h_3|^2\Big)-
\cC\Big(2 P^{(1)}(\uh)\rho_1|h_1|^2\Big) \right].
\end{split}
\end{equation}

\textbf{Case 3:} If $P_{R,l}(P) \leq P_R \leq P_{R,u}(P)$, the
optimal $(P^{(\alpha^*)}(\uh),P_R^{(\alpha^*)}(\uh))$ is given by
\begin{equation}\label{eq:fdp3}
\begin{split}
& P^{(\alpha^*)}(\uh)= \\
& \;\; \begin{cases} \text{positive root $x$ of \eqref{eq:root} if
it exists,
otherwise } 0, \\
\hspace{4.8cm} \text{if}\;\; \uh \in A; \\
\displaystyle \frac{1}{2} \left( \frac{1}{\lambda \ln
2}-\frac{1}{\rho_1 |h_1|^2} \right)^+, \hspace{1.3cm}
\text{if}\;\; \uh \in A^c
\end{cases}
\end{split}
\end{equation}
where the root $x$ is determined by the following equation
\begin{equation}\label{eq:root}
\frac{\alpha^*}{2\ln 2}\cdot \frac{1}{\frac{1}{2\rho_1|h_1|^2}+ x}
+ \frac{1-\alpha^*}{2\ln 2}\cdot
\frac{1}{\frac{1}{2\rho_3|h_3|^2}+ x} -\lambda = 0.
\end{equation}
\begin{equation}\label{eq:fdpr3}
P_R^{(\alpha^*)}(\uh)= \frac{1}{2}\left( \frac{\alpha^*}{\mu \ln
2}-\frac{1}{\rho_2 |h_2|^2} \right)^+
\end{equation}
The parameters $\lambda$ and $\mu$ are chosen to satisfy the power
constraints given in \eqref{eq:fullcg}. The parameter $\alpha^*$
is determined by the following condition:
\begin{equation}\label{eq:compare3}
\begin{split}
\mE &\left[ \cC\Big(2 P^{(\alpha^*)}(\uh)\rho_1 |h_1|^2 \Big)+
\cC\Big(2 P_R^{(\alpha^*)}(\uh)\rho_2|h_2|^2\Big)\right]\\
&= \mE_A \left[\cC\Big( 2
P^{(\alpha^*)}(\uh)\rho_3|h_3|^2\Big)\right] \\
& \hspace{4mm} +\mE_{A^c}\left[ \cC\Big(
2P^{(\alpha^*)}(\uh)\rho_1|h_1|^2\Big) \right].
\end{split}
\end{equation}
\end{small}
\rule{\linewidth}{0.3mm}
\end{quote}

%Therefore, Case 1 occurs if $P_R \ge P_{Ru}(P)$, i.e., the relay
%power is large compared to the source power. For this case, the
%optimal power allocation functions $P(\uh)$ and $P_R(\uh)$ have
%{\em water-filling} forms given by \eqref{eq:fdp1} and
%\eqref{eq:fdpr1}. Case 2 occurs if $P_r \leq P_{Rl}(P)$, i.e., the
%relay power is small compared to the source power. For this case,
%the optimal power allocation functions $P(\uh)$ and $P_R(\uh)$
%have {\em water-filling} forms given by \eqref{eq:fdp2} and
%\eqref{eq:fdpr2}.

It can be seen that in all cases the optimal power allocation
$P_R(\uh)$ for the relay node depends only on the fading gain
$h_2$ of the relay-to-destination link and it is always a
water-filling solution. However, the optimal power allocation
$P(\uh)$ for the source node in general depends on the fading
gains $h_1$ and $h_3$ corresponding to two links
(source-to-destination and source-to-relay), and it is not a
water-filling solution in general. Only in cases where $P_R$ is
large or small compared to $P$, i.e., where $P_R \ge P_{R,u}(P)$
or $P_R \leq P_{R,l}(P)$, the optimal $P(\uh)$ depends only on the
fading gain of one link and it reduces to a water-filling
solution. This is intuitive because when $P_R$ is small compared
to $P$, we should make the multiple access transmission from the
source and relay nodes to the destination node as strong as
possible, and hence the power allocation at the source node should
be based on the fading gain $h_1$ of the source-to-destination
link. When $P_R$ is large compared to $P$, we should transmit as
much information as possible from the source node to the relay
node, and hence the power allocation at the source node should be
based on the fading gain $h_3$ of the source-to-relay link.

\begin{figure}[tbhp]
\begin{center}
\begin{psfrags}
\epsfig{file=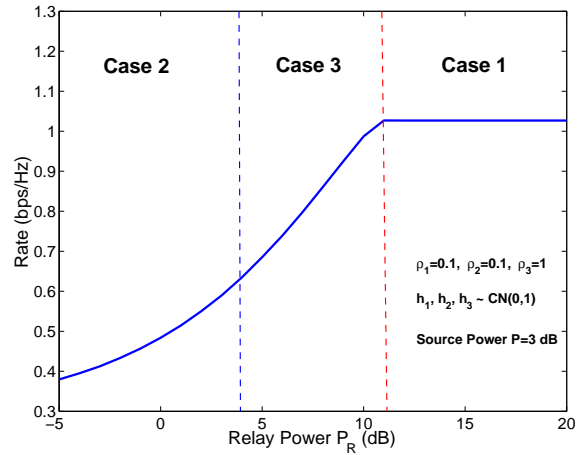,width=8cm}
\end{psfrags}
\caption{Optimal achievable rates in Scenario I}
\label{fig:gauss1}
\end{center}
\end{figure}

We now provide numerical results for a Rayleigh fading half-duplex
relay channel. We assume that the fading coefficients $h_1$, $h_2$
and $h_3$ are independent, zero-mean, unit variance, proper
complex Gaussian random variables (i.e., the amplitudes $|h_1|$,
$|h_2|$ and $|h_3|$ have a Rayleigh distribution). We further
assume $\rho_1=0.1$, $\rho_2=0.1$, and $\rho_3=1$. We assume the
power constraint at the source node is $P=3$ dB. This corresponds
to the practical environment where the relay node is close to the
source node. In Fig.~\ref{fig:gauss1}, we plot the achievable
rates for Scenario I optimized over power allocation
$(P(\uh),P_R(\uh))$. We also indicate the corresponding max-min
optimization cases to achieve these rates. It can be seen that the
achievable rate increases as the relay power increases in cases 2
and 3, and saturates when the relay power falls into case 1. This
is because in case 1 the relay power is large enough to forward
all the information decoded at the relay node to the destination
node, and the achievable rate is limited by the capacity of the
source-to-relay link.

\begin{figure}[tbhp]
\begin{center}
\begin{psfrags}
\epsfig{file=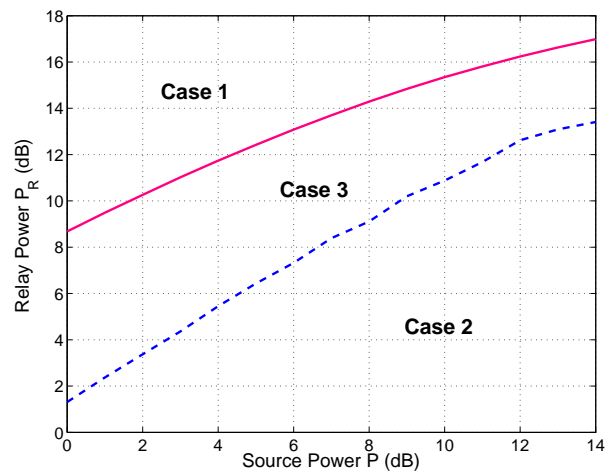,width=8cm}
\end{psfrags}
\caption{Ranges of source and relay powers with corresponding
max-min optimization cases in Scenario I} \label{fig:cases1}
\end{center}
\end{figure}

In Fig.~\ref{fig:cases1}, we plot the ranges of the source and
relay powers with their corresponding max-min optimization cases.
The solid line in the graph divides cases 1 and 3, and corresponds
to the threshold function $P_{R,u}(P)$. The dashed line divides
cases 2 and 3, and corresponds to the threshold function
$P_{R,l}(P)$. It is clear from the graph that when the relay power
is small compared to the source power, the optimal power
allocation falls into case 2, and when the relay power is large
compared to the source power, the optimal power allocation falls
into case 1. Since the achievable rate (based on the
decode-and-forward scheme) saturates in case 1, it is not useful
to increase the relay power beyond the solid line in
Fig.~\ref{fig:cases1} if the decode-and-forward scheme is adopted.
Hence the solid line $P_{R,u}$ defines the relay powers that
provide the best decode-and-forward rates under Scenario I for the
corresponding source powers.

\subsection{Scenario II: Same $\theta$ for All Channel States}

In Scenario I, $\theta$ is fixed at $1/2$; i.e., the channel
resource of time and bandwidth is equally allocated for the two
orthogonal channels. Such equal channel resource allocation may
not be optimal, and therefore we consider Scenario II, where the
channel resource allocation parameter $\theta$ needs to be
optimized jointly with power allocation. We also assume that
$\theta$ is the same for all channel states to make the system
design simple. As in Scenario I, the optimal solution of Scenario
II can also be implemented in a distributed manner at the source
and relay nodes. This is because the optimal $\theta$ depends only
on the channel statistics, not on the channel state realizations.
The power allocation at each node depends only on the channel
state of the links over which the node transmits.

We first give an achievable rate (lower bound on the capacity) and
a cut-set upper bound on the capacity. We then study the joint
channel resource and power allocations that optimize these bounds.
We also characterize the condition when the two bounds match and
establish the capacity.
\begin{proposition}
An achievable rate for the fading half-duplex relay channel
scenario II is given by
\begin{equation}\label{eq:fdlowII}
\begin{split}
& C_{\mathtt{low}} = \max_{0\leq \theta \leq 1, (P(\uh),P_R(\uh))
\in \cG} \\
& \min \Bigg \{ \mE \left[2\theta \cC\left(\frac{ P(\uh)\rho_1
|h_1|^2}{\theta} \right)+ 2\bartheta \cC\left(
\frac{P_R(\uh)\rho_2|h_2|^2}{\bartheta}\right) \right] ,   \\
& \mE_A \left[ 2\theta
\cC\left(\frac{P(\uh)\rho_3|h_3|^2}{\theta}\right)\right] +
\mE_{A^c}\left[ 2\theta\cC\left(
\frac{P(\uh)\rho_1|h_1|^2}{\theta}\right) \right] \Bigg \}
\end{split}
\end{equation}
where $\bartheta=1-\theta$. An upper bound on the capacity is
given by
\begin{equation}\label{eq:fdupII}
\begin{split}
& C_{\mathtt{up}} = \max_{0 \leq \theta \leq
1,(P(\uh),P_R(\uh))\in \cG} \\
& \min \Bigg \{ \mE \left[ 2\theta \cC\left(\frac{ P(\uh)\rho_1
|h_1|^2}{\theta} \right)+ 2\bartheta \cC\left(
\frac{P_R(\uh)\rho_2|h_2|^2}{\bartheta}\right) \right],   \\
&  \hspace{1cm} \mE\left[ 2\theta
\cC\left(\frac{P(\uh)(\rho_3|h_3|^2+\rho_1
|h_1|^2)}{\theta}\right) \right]  \Bigg \} .
\end{split}
\end{equation}
\end{proposition}

We provide the optimal channel resource and power allocations
$(\theta,P(\uh),P_R(\uh))$ that solve \eqref{eq:fdlowII} in the
following. The proof of optimality is relegated to Appendix
\ref{app:scenarioII}.

\begin{quote}
\textbf{Optimal resource allocation that maximizes the lower bound
\eqref{eq:fdlowII}:}

\vspace{-2mm}
 \rule{\linewidth}{0.3mm}

\begin{small}
\textbf{Case 1:} This case is included in case 3 with the
parameter $\alpha$ being allowed to take the value of $0$.

\textbf{Case 2:} The optimal
$(\theta^{(1)},P^{(1)}(\uh),P^{(1)}_R(\uh) )$ can be obtained by
the following iterative algorithm. For a given $\theta$, the power
allocation $(P(\uh),P_R(\uh) )$ are given by
\begin{equation}\label{eq:fdp2II}
P(\uh)= \theta\left( \frac{1}{\lambda \ln 2}-\frac{1}{\rho_1
|h_1|^2} \right)^+
\end{equation}
\begin{equation}\label{eq:fdpr2II}
P_R(\uh)= \bartheta\left( \frac{1}{\mu \ln
2}-\frac{1}{\rho_2|h_2|^2} \right)^+ ,
\end{equation}
where $\lambda$ and $\mu$ are chosen to satisfy the power
constraints. For a given $(P(\uh),P_R(\uh) )$, the value of
$\theta$ is given by the root of the following equation:
\begin{equation}\label{eq:fdtheta2II}
\begin{split}
2\mE & \left[\cC\left(\frac{P(\uh)\rho_1 |h_1|^2}{\theta} \right)
- \cC\left( \frac{P_R(\uh)\rho_2|h_2|^2}{\bartheta}\right)\right]
\\
& = \frac{1}{\ln 2}\mE \left[ \frac{ P(\uh) \rho_1
|h_1|^2}{\theta+P(\uh) \rho_1 |h_1|^2}- \frac{ P_R(\uh) \rho_2
|h_2|^2}{\bartheta+P_R(\uh) \rho_2 |h_2|^2}\right].
\end{split}
\end{equation}
The resource allocation $(\theta,P(\uh),P_R(\uh))$ obtained
iteratively from \eqref{eq:fdp2II},\eqref{eq:fdpr2II}, and
\eqref{eq:fdtheta2II} converges to the optimal
$(\theta^{(1)},P^{(1)}(\uh),P^{(1)}_R(\uh))$.

This case happens if the following condition is satisfied
\begin{equation}\label{eq:compare2II}
\begin{split}
& \bartheta^{(1)} \mE \left[\cC\left(\frac{
P_R^{(1)}(\uh)\rho_2|h_2|^2}{\bartheta^{(1)}}\right) \right]
\\
& \pp \leq \theta^{(1)} \mE_A \Bigg[\cC\left( \frac{
P^{(1)}(\uh)\rho_3|h_3|^2}{\theta^{(1)}}\right) \\
& \hspace{2.5cm} - \cC\left(\frac{
P^{(1)}(\uh)\rho_1|h_1|^2}{\theta^{(1)}}\right)\Bigg].
\end{split}
\end{equation}

\textbf{Case 3:} The optimal
$(\theta^{(\alpha^*)},P^{(\alpha^*)}(\uh),P^{(\alpha^*)}_R(\uh))$
can be obtained by the following iterative algorithm. For a given
$\theta$, the power allocation $(P(\uh),P_R(\uh) )$ is given by
\begin{equation}\label{eq:fdp3II}
P(\uh)=
\begin{cases}
\text{positive root $x$ of \eqref{eq:rootII} if it exists,
otherwise } 0, \\
\hspace{4.7cm} \text{if}\;\; \uh \in A, \\
\displaystyle \theta\left( \frac{1}{\lambda \ln 2}-\frac{1}{\rho_1
|h_1|^2} \right)^+ \\
\hspace{4.6cm} \text{if}\;\; \uh \in A^c;
\end{cases}
\end{equation}
where the root $x$ is determined by the following equation:
\begin{equation}\label{eq:rootII}
\frac{\alpha^* \theta}{\ln 2} \cdot
\frac{1}{\frac{\theta}{\rho_1|h_1|^2}+ x} +
\frac{(1-\alpha^*)\theta}{\ln 2}\cdot
\frac{1}{\frac{\theta}{\rho_3|h_3|^2}+ x} -\lambda = 0.
\end{equation}
\begin{equation}\label{eq:fdpr3II}
P_R(\uh)= \bartheta\left( \frac{\alpha^*}{\mu \ln
2}-\frac{1}{\rho_2 |h_2|^2} \right)^+ .
\end{equation}
The parameters $\lambda$ and $\mu$ are chosen to satisfy the power
constraints.

For a given $(P(\uh),P_R(\uh))$, the value of $\theta$ is the root
of the following equation:
\begin{equation} \label{eq:fdtheta3II}
\begin{split}
& \textstyle 2\alpha^* \mE_A \left[\cC\left(\frac{P(\uh)\rho_1
|h_1|^2}{\theta} \right)\right] - \frac{\alpha^*}{\ln 2}\mE_A
\left[ \frac{ P(\uh)
\rho_1|h_1|^2}{\theta+P(\uh) \rho_1 |h_1|^2}\right]\\
& \textstyle -2\alpha^* \mE \left[\cC\left(
\frac{P_R(\uh)\rho_2|h_2|^2}{\bartheta}\right)\right] +
\frac{\alpha^*}{\ln 2}\mE\left[\frac{ P_R(\uh) \rho_2
|h_2|^2}{\bartheta+P_R(\uh) \rho_2
|h_2|^2} \right] \\
& \textstyle + 2\mE_{A^c}\left[ \cC\left(\frac{P(\uh)\rho_1
|h_1|^2}{\theta} \right)\right]- \frac{1}{\ln 2}\mE_{A^c}\left[
\frac{ P(\uh) \rho_1 |h_1|^2}{\theta+P(\uh) \rho_1 |h_1|^2}
\right]
\\
& \textstyle  + 2(1-\alpha^*) \mE_A
\left[\cC\left(\frac{P(\uh)\rho_3 |h_3|^2}{\theta} \right)\right]
\\
& \textstyle-\frac{1-\alpha^*}{\ln 2}\mE_A \left[\frac{ P(\uh)
\rho_3
|h_3|^2}{\theta+P(\uh) \rho_3 |h_3|^2}\right]  \\
& \spp =0
\end{split}
\end{equation}
The resource allocation $(\theta,P(\uh),P_R(\uh))$ obtained
iteratively from \eqref{eq:fdp3II},\eqref{eq:fdpr3II}, and
\eqref{eq:fdtheta3II} converges to the optimal
$(\theta^{(\alpha^*)},P^{(\alpha^*)}(\uh),P^{(\alpha^*)}_R(\uh))$.

Finally, the parameter $\alpha^*$ is determined by the condition
\begin{equation}\label{eq:compare3II}
\begin{split}
\mE \Bigg[ & \theta^{(\alpha^*)} \cC\left(\frac{
P^{(\alpha^*)}(\uh)\rho_1 |h_1|^2}{\theta^{(\alpha^*)}} \right)
\\
& + \bartheta^{(\alpha^*)} \cC\left(
\frac{P_R^{(\alpha^*)}(\uh)\rho_2|h_2|^2}{\bartheta^{(\alpha^*)}}\right) \Bigg],   \\
& \pp = \mE_A \left[ \theta^{(\alpha^*)}
\cC\left(\frac{P^{(\alpha^*)}(\uh)\rho_3|h_3|^2}{\theta^{(\alpha^*)}}\right)\right]
\\
& \hspace{1.4cm}+ \mE_{A^c}\left[ \theta^{(\alpha^*)}\cC\left(
\frac{P^{(\alpha^*)}(\uh)\rho_1|h_1|^2}{\theta^{(\alpha^*)}}\right)
\right] .
\end{split}
\end{equation}
\end{small}
\rule{\linewidth}{0.3mm}
\end{quote}

The optimization for the upper bound \eqref{eq:fdupII} can be
performed in a similar manner, and is not presented in this paper.
In general, the lower bound \eqref{eq:fdlowII} and the upper bound
\eqref{eq:fdupII} do not match. In the following theorem, we
characterize the condition under which the two bounds match and
hence yield the capacity of this channel.
\begin{theorem}\label{th:halfIIcapa}
For the fading half-duplex relay channel Scenario II, if the
channel statistics and the power constraints satisfy the condition
\eqref{eq:compare2II}, then the capacity is given by
\begin{equation}\label{eq:halfIIcapa}
\begin{split}
C = \mE \Bigg[ & 2\theta^{(1)} \cC\left(\frac{ P^{(1)}(\uh)\rho_1
|h_1|^2}{\theta^{(1)}} \right) \\
& + 2\bartheta^{(1)} \cC\left(
\frac{P^{(1)}_R(\uh)\rho_2|h_2|^2}{\bartheta}\right) \Bigg]
\end{split}
\end{equation}
where the capacity achieving resource allocation
$(\theta^{(1)},P^{(1)}(\uh),P_R^{(1)}(\uh))$ can be obtained
iteratively from \eqref{eq:fdp2II}, \eqref{eq:fdpr2II} and
\eqref{eq:fdtheta2II}.
\end{theorem}

The proof of Theorem \ref{th:halfIIcapa} is similar to the proof
of Theorem \ref{th:fullacapa}, and is hence omitted.
\begin{remark}
The capacity in Theorem \ref{th:halfIIcapa} refers to the largest
rate under Scenario II that can be achieved over all possible
channel resource allocation parameters $\theta$ and over all
possible power allocation rules $(P(\uh),P_R(\uh))$.
\end{remark}

The condition given in Theorem \ref{th:halfIIcapa} tends to be
satisfied either when the relay power $P_R$ is small compared to
the source power $P$, or when the relay is much closer to the
source than to the destination.

\begin{figure}[tbhp]
\begin{center}
\begin{psfrags}
\epsfig{file=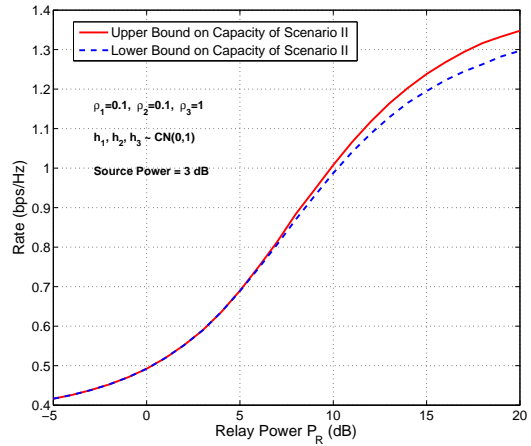,width=8cm}
\end{psfrags}
\caption{Lower and upper bounds on capacity of Scenario II}
\label{fig:lowup2}
\end{center}
\end{figure}

In Fig.~\ref{fig:lowup2}, we plot the lower and upper bounds on
the capacity of Scenario II for the same Rayleigh fading relay
channel as in Fig.~\ref{fig:gauss1}. Both bounds are optimized
over $(\theta,P(\uh),P_R(\uh))$. It can be seen from
Fig.~\ref{fig:lowup2} that when the relay power is less than a
threshold ($4$ dB), the two bounds match and determine the
capacity of Scenario II. This demonstrates our capacity result in
Theorem \ref{th:halfIIcapa} and the condition when the lower and
upper bounds match. Fig.~\ref{fig:lowup2} also shows that the gap
between the lower and upper bounds is small even when the relay
power is large.

\begin{figure}[tbhp]
\begin{center}
\begin{psfrags}
\epsfig{file=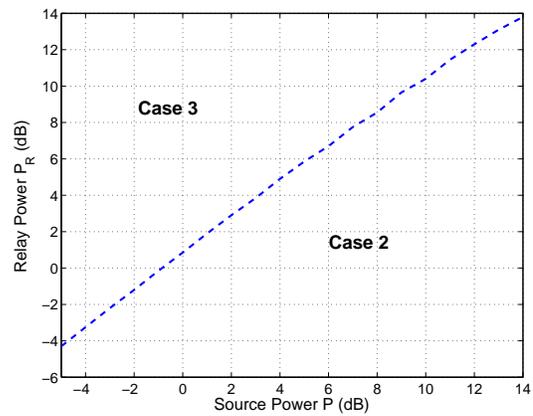,width=7.cm}
\end{psfrags}
\caption{Ranges of source and relay powers with corresponding
max-min optimization cases in Scenario II} \label{fig:cases2}
\end{center}
\end{figure}

\begin{figure}[tbhp]
\begin{center}
\begin{psfrags}
\epsfig{file=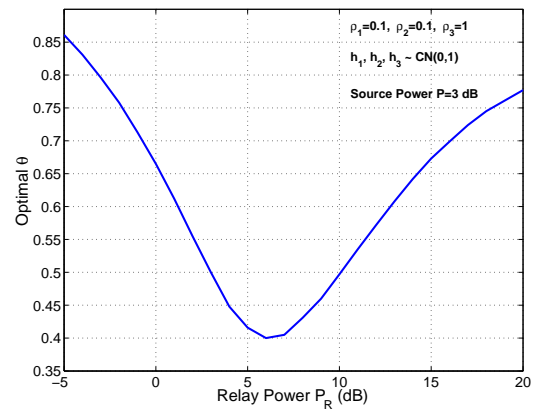,width=7cm}
\end{psfrags}
\caption{Optimal $\theta$ as a function of relay power in Scenario
II} \label{fig:theta2}
\end{center}
\end{figure}

In Fig.~\ref{fig:cases2}, we plot the ranges of the source and
relay powers with their corresponding max-min optimization cases.
The dashed line in the graph divides cases 2 and 3. Similar to
Fig.~\ref{fig:cases1}, the optimal power allocation falls into
case 2 when the relay power is small compared to the source power.
However, we see that Fig.~\ref{fig:cases2} deviates from
Fig.~\ref{fig:cases1} in that case 1 (where the achievable rate
saturates) is missing in Scenario II. This explains why the
achievable rate under Scenario II continues to increase beyond the
point where the rate under Scenario I saturates (see
Fig.~\ref{fig:gauss123} in Section \ref{sec:scenarioIII}).

%This also explains the reason that the achievable rate of Scenario
%II continues to increase when the achievable rate of Scenario I
%becomes saturate.

In Fig.~\ref{fig:theta2}, we plot the optimal value of $\theta$ as
a function of the relay power, and observe that it is not a
monotonic function. When the relay power is small, as the relay
power increases, the optimal $\theta$ decreases so that more of
the channel resource is assigned to the relay-to-destination link
to make more use of the relay node. When the relay power is large,
as the relay power increases, the optimal $\theta$ increases. This
is because the relay power is now large enough to forward all the
information decoded at the relay node to the destination node even
with a small amount of the channel resource, and hence more of the
channel resource is needed for the source node to transmit more
information to the relay node. This behavior of the optimal
$\theta$ is similar to that of the Gaussian half-duplex relay
channel studied in \cite{Liang05it}.

\subsection{Scenario III: $\theta$ Changes with Channel
States}\label{sec:scenarioIII}

In Scenario II, the parameter $\theta$ is required to be the same
for all channel states, and only the power allocations are
dynamically adjusted according to the instantaneous channel state.
In this subsection, we study Scenario III, where $\theta$ is also
allowed to change with the channel state realizations, and
$\theta(\uh)$ is optimized jointly with power allocation
$(P(\uh),P_R(\uh))$. However, for the source and relay nodes to
decide $\theta(\uh)$ for each channel state, each node needs to
know the channel realizations on all transmission links. This
makes the system design more complex, and not as practical as
Scenario II. We include the analysis of the resource allocation
for this scenario mainly for the sake of completeness.

%For Scenario III, we give the lower and upper bounds on the
%capacity and optimize these bounds over
%$(\theta(\uh),P(\uh),P_R(\uh))$ similarly as for Scenario II. We
%also characterize the condition for the lower bound to achieve the
%upper bound, and thus obtain the capacity for this scenario when
%this condition is satisfied.
\begin{proposition}
An achievable rate for the fading half-duplex relay channel
Scenario III is given by:
\begin{equation}\label{eq:fdlowIII}
\begin{split}
& C_{\mathtt{low}} = \max_{ \begin{array}{l} \scriptstyle 0 \leq
\theta(\uh) \leq 1, \\ \scriptstyle (P(\uh),P_R(\uh))\in \cG
\end{array}} \min \\
& \textstyle \Bigg \{ \mE \left[2\theta(\uh)\cC\left(\frac{
P(\uh)\rho_1 |h_1|^2}{\theta(\uh)} \right)\right]+\mE \left[2
\bartheta(\uh)\cC\left(
\frac{P_R(\uh)\rho_2|h_2|^2}{\bartheta(\uh)}\right)\right], \\
& \textstyle \hspace{0.2cm} \mE_A\left[2\theta(\uh)
\cC\left(\frac{P(\uh)\rho_3|h_3|^2}{\theta(\uh)}\right)\right] +
\mE_{A^c}\left[2
\theta(\uh)\cC\Big(\frac{P(\uh)\rho_1|h_1|^2}{\theta(\uh)}\Big)
\right] \Bigg \}
\end{split}
\end{equation}

An upper bound on the capacity is given by
\begin{equation}\label{eq:fdupIII}
\begin{split}
& C_{\mathtt{up}} = \max_{ \begin{array}{l} \scriptstyle 0 \leq
\theta(\uh) \leq 1, \\ \scriptstyle (P(\uh),P_R(\uh))\in \cG
\end{array}} \min \\
& \textstyle \Bigg \{\mE \left[2\theta(\uh) \cC\left(\frac{
P(\uh)\rho_1 |h_1|^2}{\theta(\uh)} \right)\right]+ \mE
\left[2\bartheta(\uh)\cC\left(
\frac{P_R(\uh)\rho_2|h_2|^2}{\bartheta(\uh)}\right)\right],   \\
& \textstyle \hspace{3mm}\mE\left[2\theta(\uh)
\cC\left(\frac{P(\uh)(\rho_3|h_3|^2+\rho_1
|h_1|^2)}{\theta(\uh)}\right) \right]  \Bigg \} .
\end{split}
\end{equation}
\end{proposition}

The optimal resource allocation $(\theta(\uh),P(\uh),P_R(\uh))$
that achieves the maximum of the lower bound \eqref{eq:fdlowIII}
is given in the following. The proof of optimality is relegated to
Appendix \ref{app:scenarioIII}.

\begin{quote}
\textbf{Optimal resource allocation that maximizes the lower bound
\eqref{eq:fdlowIII}:}

\vspace{-2mm}
 \rule{\linewidth}{0.3mm}

\begin{small}
\textbf{Case 1:} The optimal resource allocation
$(\theta^{(0)}(\uh),P^{(0)}(\uh),P_R^{(0)}(\uh))$ is given by
\begin{equation}\label{eq:fdp1III}
P^{(0)}(\uh)=
\begin{cases}
\displaystyle \left( \frac{1}{\lambda \ln 2}-\frac{1}{\rho_3
|h_3|^2} \right)^+ , &  \spp \text{if}\;\; \uh \in A,  \\
\displaystyle \left( \frac{1}{\lambda \ln 2}-\frac{1}{\rho_1
|h_1|^2} \right)^+ , &  \spp \text{if}\;\; \uh \in A^c;
\end{cases}
\end{equation}
\begin{equation}\label{eq:fdpr1III}
P^{(0)}_R(\uh)=
\begin{cases}
\displaystyle \left( \frac{1}{\mu \ln 2}-\frac{1}{\rho_2|h_2|^2}
\right)^+ , \;\; &
\text{if } P^{(0)}(\uh)=0; \\
\;\; 0, &  \text{if } P^{(0)}(\uh)> 0,
\end{cases}
\end{equation}
where $\lambda$ and $\mu$ are chosen to satisfy the power
constraint given in \eqref{eq:fullcg}.
\begin{equation}\label{eq:fdtheta1III}
\theta^{(0)}(\uh)=
\begin{cases}
1, \pp \mbox{if }\; P^{(0)}(\uh) > 0 ; \\
0, \pp \mbox{if }\; P^{(0)}(\uh) = 0 .
\end{cases}
\end{equation}

For case 1 to happen,
$(\theta^{(0)}(\uh),P^{(0)}(\uh),P^{(0)}_R(\uh))$ needs to satisfy
the following condition:
\begin{equation}\label{eq:compare1III}
\begin{split}
& \mE\left[ \cC\left(P^{(0)}_R(\uh)\rho_2|h_2|^2\right) \right]
\\
& \spp \ge  \mE_A \left[
\cC\left(P^{(0)}(\uh)\rho_3|h_3|^2\right)-
\cC\left(P^{(0)}(\uh)\rho_1|h_1|^2\right) \right].
\end{split}
\end{equation}

\textbf{Case 2:} The optimal
$(\theta^{(1)}(\uh),P^{(1)}(\uh),P^{(1)}_R(\uh))$ can be
determined by the following iterative algorithm. For a given
$\theta(\uh)$, the power allocation $(P(\uh),P_R(\uh))$ is given
by
\begin{equation}\label{eq:fdp2III}
P(\uh)= \theta(\uh)\left( \frac{1}{\lambda \ln 2}-\frac{1}{\rho_1
|h_1|^2} \right)^+
\end{equation}
\begin{equation}\label{eq:fdpr2III}
P_R(\uh)= \bartheta(\uh)\left( \frac{1}{\mu \ln
2}-\frac{1}{\rho_2|h_2|^2} \right)^+
\end{equation}
where $\lambda$ and $\mu$ are chosen to satisfy the power
constraints given in \eqref{eq:fullcg}. For a given
$(P(\uh),P_R(\uh))$, the channel resource allocation $\theta(\uh)$
is the root of the following equation:
\begin{equation} \label{eq:fdtheta2III}
\begin{split}
& 2\cC\left(\frac{P(\uh)\rho_1 |h_1|^2}{\theta(\uh)} \right)
-2\cC\left( \frac{P_R(\uh)\rho_2|h_2|^2}{\bartheta(\uh)}\right)
\\
&- \frac{1}{\ln 2} \frac{ P(\uh) \rho_1
|h_1|^2}{\theta(\uh)+P(\uh) \rho_1 |h_1|^2}+ \frac{1}{\ln 2}
\frac{ P_R(\uh) \rho_2 |h_2|^2}{\bartheta(\uh)+P_R(\uh) \rho_2
|h_2|^2} \\
& \spp = 0.
\end{split}
\end{equation}

For case 2 to happen,
$(\theta^{(1)}(\uh),P^{(1)}(\uh),P^{(1)}_R(\uh))$ needs to satisfy
the following condition:
\begin{equation}\label{eq:compare2III}
\begin{split}
\mE & \left[\bartheta^{(1)}(\uh)\cC\left(\frac{
P^{(1)}_R(\uh)\rho_2|h_2|^2}{\bartheta^{(1)}(\uh)}\right)\right]  \\
& \leq  \mE_A \Bigg[\theta^{(1)}(\uh)\cC\left( \frac{
P^{(1)}(\uh)\rho_3|h_3|^2}{\theta^{(1)}(\uh)}\right) \\
& \hspace{1cm} -\theta^{(1)}(\uh)\cC\left(\frac{
P^{(1)}(\uh)\rho_1|h_1|^2}{\theta^{(1)}(\uh)}\right) \Bigg]
\end{split}
\end{equation}

\textbf{Case 3:} The optimal
$(\theta^{(\alpha^*)}(\uh),P^{(\alpha^*)}(\uh),P_R^{(\alpha^*)}(\uh))$
can be obtained by the following iterative algorithm. For a given
$\theta(\uh)$, the power allocation $(P(\uh),P_R(\uh))$ is given
by
\begin{equation}\label{eq:fdp3III}
P(\uh)=
\begin{cases}
\text{positive root $x$ of \eqref{eq:rootIII} if it exists,
otherwise } 0, \\
\hspace{4.7cm} \text{if}\;\; \uh \in A, \\
\displaystyle \theta(\uh) \left( \frac{1}{\lambda \ln
2}-\frac{1}{\rho_1 |h_1|^2} \right)^+, \hspace{0.8cm}
\text{if}\;\; \uh \in A^c;
\end{cases}
\end{equation}
where the root $x$ is determined by the following equation:
\begin{equation}\label{eq:rootIII}
\frac{\alpha^* \theta(\uh)}{\ln 2}
\frac{1}{\frac{\theta(\uh)}{\rho_1|h_1|^2}+ x} +
\frac{(1-\alpha^*)\theta(\uh)}{\ln 2}
\frac{1}{\frac{\theta(\uh)}{\rho_3|h_3|^2}+ x} -\lambda = 0.
\end{equation}
\begin{equation}\label{eq:fdpr3III}
P_R(\uh)= \bartheta(\uh)\left( \frac{\alpha^*}{\mu \ln
2}-\frac{1}{\rho_2 |h_2|^2} \right)^+
\end{equation}
where the parameters $\lambda$ and $\mu$ are chosen to satisfy the
power constraints \eqref{eq:fullcg}.

For a given $(P(\uh),P_R(\uh))$, the channel resource allocation
$\theta(\uh)$ is determined by
\begin{equation}\label{eq:theta3III}
\begin{split}
& \text{If } \uh \in A, \\
& 2\alpha^*\cC\left(\frac{P(\uh)\rho_1
|h_1|^2}{\theta(\uh)} \right) - \frac{\alpha^*}{\ln 2} \frac{
P(\uh) \rho_1
|h_1|^2}{\theta(\uh)+P(\uh) \rho_1 |h_1|^2} \\
& -2\alpha^* \cC\left(
\frac{P_R(\uh)\rho_2|h_2|^2}{\bartheta(\uh)}\right) +
\frac{\alpha^*}{\ln 2}\frac{ P_R(\uh) \rho_2
|h_2|^2}{\bartheta(\uh)+P_R(\uh) \rho_2
|h_2|^2} \\
& + 2(1-\alpha*) \cC\left(\frac{P(\uh)\rho_3 |h_3|^2}{\theta(\uh)}
\right) \\
& -\frac{1-\alpha*}{\ln 2} \frac{ P(\uh) \rho_3
|h_3|^2}{\theta(\uh)+P(\uh) \rho_3 |h_3|^2}  \\
& \spp =0 ;
\end{split}
\end{equation}
\begin{equation}
\begin{split}
& \text{If } \uh \in A^c, \\
& -2\alpha^* \cC\left(
\frac{P_R(\uh)\rho_2|h_2|^2}{\bartheta(\uh)}\right) +
\frac{\alpha^*}{\ln 2}\frac{ P_R(\uh) \rho_2
|h_2|^2}{\bartheta(\uh)+P_R(\uh) \rho_2
|h_2|^2} \hspace{1.7cm} \\
& + 2\cC\left(\frac{P(\uh)\rho_1 |h_1|^2}{\theta(\uh)} \right)
-\frac{1}{\ln 2} \frac{ P(\uh) \rho_1
|h_1|^2}{\theta(\uh)+P(\uh) \rho_1 |h_1|^2} \nonumber  \\
& \spp = 0.
\end{split}
\end{equation}
The parameter $\alpha^*$ is determined by the condition
\begin{equation}
\begin{split}
& \mE \left[\theta^{(\alpha^*)}(\uh) \cC\left(\frac{
P^{(\alpha^*)}(\uh)\rho_1 |h_1|^2}{\theta^{(\alpha^*)}(\uh)}
\right)\right] \\
& + \mE \left[\bartheta^{(\alpha^*)}(\uh)\cC\left(
\frac{P^{(\alpha^*)}_R(\uh)\rho_2|h_2|^2}{\bartheta^{(\alpha^*)}(\uh)}\right)\right],   \\
& \pp = \mE_A\left[\theta^{(\alpha^*)}(\uh)
\cC\left(\frac{P^{(\alpha^*)}(\uh)\rho_3|h_3|^2}{\theta^{(\alpha^*)}(\uh)}\right)\right]
\\
& \hspace{1.4cm} +\mE_{A^c}\left[
\theta^{(\alpha^*)}(\uh)\cC\Big(\frac{P^{(\alpha^*)}(\uh)\rho_1|h_1|^2}{\theta^{(\alpha^*)}(\uh)}\Big)
\right].
\end{split}
\end{equation}
\end{small}
\rule{\linewidth}{0.3mm}
\end{quote}

The optimization for the upper bound \eqref{eq:fdupIII} can be
performed using steps that are similar to those for the lower
bound. In general, the lower bound \eqref{eq:fdlowIII} and the
upper bound \eqref{eq:fdupIII} do not match. However, we show that
if the channel statistics and the power constraints satisfy the
following condition, the two bounds match and hence we obtain the
capacity for this channel.
\begin{theorem}\label{th:halfIIIcapa}
For the fading half-duplex relay channel Scenario III, if the
channel statistics and the power constraints satisfy the condition
\eqref{eq:compare2III}, then the capacity is given by
\begin{equation}\label{eq:halfIIIcapa}
\begin{split}
C= & \mE \left[2\theta^{(1)}(\uh)\cC\left(\frac{
P^{(1)}(\uh)\rho_1 |h_1|^2}{\theta^{(1)}(\uh)} \right)\right]
\\
& +\mE \left[2 \bartheta^{(1)}(\uh)\cC\left(
\frac{P_R^{(1)}(\uh)\rho_2|h_2|^2}{\bartheta^{(1)}(\uh)}\right)\right]
\end{split}
\end{equation}
where the capacity achieving resource allocation
$(\theta^{(1)}(\uh),P^{(1)}(\uh),P_R^{(1)}(\uh))$ can be obtained
iteratively from \eqref{eq:fdp2III}, \eqref{eq:fdpr2III} and
\eqref{eq:fdtheta2III}.
\end{theorem}

The proof of Theorem \ref{th:halfIIIcapa} is similar to that of
Theorem \ref{th:fullacapa}, and is omitted.
\begin{remark}
The capacity in Theorem \ref{th:halfIIIcapa} refers to the largest
rate under Scenario III that can be achieved over all possible
channel resource allocation $\theta(\uh)$ and power allocation
$(P(\uh),P_R(\uh))$.
\end{remark}

The condition given in Theorem \ref{th:halfIIIcapa} is similar to
that in Theorem \ref{th:halfIIcapa} for Scenario II, and these
conditions tend to be satisfied either when the relay power $P_R$
is small compared to the source power $P$, or when the relay node
is much closer to the source node than to the destination node.

\begin{figure}[tbhp]
\begin{center}
\begin{psfrags}
\epsfig{file=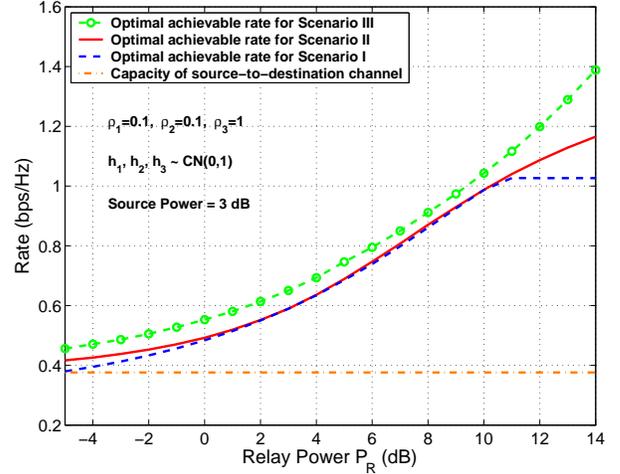,width=8cm}
\end{psfrags}
\caption{Comparison of achievable rates with optimal resource
allocations for Scenarios I, II, and III} \label{fig:gauss123}
\end{center}
\end{figure}

In Fig.~\ref{fig:gauss123}, we plot the achievable rates under
Scenario III optimized over $(\theta(\uh),P(\uh),P_R(\uh))$ for
the same Rayleigh fading relay channel as in Fig.~\ref{fig:gauss1}
and Fig.~\ref{fig:lowup2}. We compare these rates with the
achievable rates under Scenario I optimized over
$(P(\uh),P_R(\uh))$ and Scenario II optimized over
$(\theta,P(\uh),P_R(\uh))$, and with the capacity of the direct
link from the source node to the destination node. It is clear
from the graph that employing the relay node greatly improves the
performance of the source-to-destination channel.
Fig.~\ref{fig:gauss123} shows that the achievable rate under
scenario II is larger than the achievable rate under Scenario I,
particularly when the relay power is large and the achievable rate
under scenario I saturates. This demonstrates that using a jointly
optimal channel resource allocation parameter $\theta$ helps to
improve the achievable rate. As we have commented for
Fig.~\ref{fig:cases2}, Scenario II does not have case 1, and hence
the achievable rate under Scenario II continues to increase when
the achievable rate under Scenario I saturates in case 1.
Furthermore, Scenario III has larger achievable rates than
Scenario II because $\theta(\uh)$ can be dynamically changed based
on the instantaneous channel state information.
%Since Scenario III
%also has case 1, which means the achievable rate saturates when
%the relay power is beyond a certain threshold. This rate
%saturation does not appear in the graph because it occurs only at
%very high relay powers.

We note that finding an optimal resource allocation for Rayleigh
fading relay channels is a high dimensional optimization problem,
particularly in Scenario III where the optimization is jointly
over $(\theta(\uh),P(\uh),P_R(\uh))$. Although the problem is
convex, the standard convex programming techniques may converge
slowly. However, since we have obtained the analytical structures
of the optimal solutions, our numerical algorithm converges
extremely fast and takes only a few iterations.

\section{Concluding Remarks}\label{sec:conclusion}

We have studied capacity bounds for the parallel relay channel and
its special case of the fading relay channel. We have established
capacity theorems for several classes of channels including the
parallel relay channel with degraded subchannels and its Gaussian
case, the full-duplex relay channel that satisfies certain
conditions in asynchronized case, and the half-duplex relay
channel that satisfies certain conditions.

We have studied resource allocation for the Gaussian parallel
relay channel with degraded subchannels and the fading relay
channel under both full-duplex and half-duplex models. Our study
of resource allocation is different from previous work on this
topic in that we make the more practical assumption that the
source and relay nodes are subject to separate power constraints
rather than a total power constraint. We have shown that optimal
resource allocation under this assumption may take three different
forms depending on the channel statistics and values of the power
constraints.

Finally, we note that the resource allocation problem we have
considered falls under a class of {\em max-min} problems and we
have provided a technique for solving such max-min problems. It is
known that the achievable rates of relay channels when relay nodes
use the decode-and-forward scheme are usually expressed by {\it
max-min} forms. Our technique certainly applies to optimization
problems arising in these contexts. In particular, our technique
has been applied to study orthogonal relay broadcast channels in
\cite{Liang05spawc}, and can be used to study more general classes
of relay networks with fading links.

%%%%%%%
%
%
\appendices
\section{Proof of Resource Allocation that Maximizes $C_{\mathtt{low}}$ \eqref{eq:fdlowI} for Scenario I}\label{app:scenarioI}

We first let $R_1(P(\uh),P_R(\uh))$ and $R_2(P(\uh),P_R(\uh))$
denote the two terms over which the minimization in
\eqref{eq:fdlowI} is taken. We can then express \eqref{eq:fdlowI}
in the following compact form:
\begin{equation}\label{eq:achieveI}
\begin{split}
C_{\mathtt{low}} = & \max_{(P(\uh),P_R(\uh)) \in \cG} \\
&  \min \left \{ R_1(P(\uh),P_R(\uh)),\; R_2(P(\uh),P_R(\uh))
\right \}.
\end{split}
\end{equation}

We apply Proposition \ref{prop:generule} to derive the optimal
power allocation rule, which falls into the following three cases.

\textbf{Case 1:} $\alpha^*=0$, and $(P^{(0)}(\uh),P^{(0)}_R(\uh))$
is an optimal power allocation, which needs to satisfy the
condition
\begin{equation}\label{eq:fdcon1I}
R_1(P^{(0)}(\uh),P^{(0)}_R(\uh)) \ge R_2
(P^{(0)}(\uh),P^{(0)}_R(\uh)).
\end{equation}
By definition, $(P^{(0)}(\uh),P^{(0)}_R(\uh))$ maximizes
\begin{equation}
R(0,P(\uh),P_R(\uh))=R_2(P(\uh),P_R(\uh)).
\end{equation}
The optimal $P^{(0)}(\uh)$ given in \eqref{eq:fdp1} follows easily
from the KKT condition. For case 1 to happen,
$(P^{(0)}(\uh),P^{(0)}_R(\uh))$ needs to satisfy the condition
\eqref{eq:fdcon1I}. It is clear that $R_2(P(\uh),P_R(\uh))$
depends only on $P(\uh)$. The term $R_1(P(\uh),P_R(\uh))$ depends
on both $P(\uh)$ and $P_R(\uh)$. To characterize the most general
condition for case 1 to happen, $P_R(\uh)$ needs to maximize
$R_1(P(\uh),P_R(\uh))$. Such $P_R^{(0)}(\uh)$ can be obtained by
the KKT condition and is given in \eqref{eq:fdpr1}. The condition
\eqref{eq:compare1} follows from equality of condition
\eqref{eq:fdcon1I}.

\textbf{Case 2:} $\alpha^*=1$, and $(P^{(1)}(\uh),P^{(1)}_R(\uh))$
is an optimal power allocation, which needs to satisfy the
condition
\begin{equation}\label{eq:fdcon2I}
R_1(P^{(1)}(\uh),P^{(1)}_R(\uh)) \leq
R_2(P^{(1)}(\uh),P^{(1)}_R(\uh)).
\end{equation}

The optimal $(P^{(1)}(\uh),P^{(1)}_R(\uh))$ that maximizes
\begin{equation}
R(1,P(\uh),P_R(\uh))=R_1(P(\uh),P_R(\uh))
\end{equation}
can be easily obtained by the KKT condition, and are given in
\eqref{eq:fdp2} and \eqref{eq:fdpr2}. The condition
\eqref{eq:compare2} follows from equality of condition
\eqref{eq:fdcon2I}.

\textbf{Case 3:} $0 < \alpha^* < 1$, and
$(P^{(\alpha^*)}(\uh),P^{(\alpha^*)}_R(\uh))$ is an optimal power
allocation, where $\alpha^*$ is determined by the following
condition
\begin{equation}\label{eq:fdcon3I}
R_1(P^{(\alpha^*)}(\uh),P^{(\alpha^*)}_R(\uh)) = R_2
(P^{(\alpha^*)}(\uh),P^{(\alpha^*)}_R(\uh)).
\end{equation}

We need to derive $(P^{(\alpha^*)}(\uh),P^{(\alpha^*)}_R(\uh))$
that maximizes
\begin{equation}
\begin{split}
& R(\alpha^*,P(\uh),P_R(\uh)) \\
& \spp =\alpha^*
R_1(P(\uh),P_R(\uh))+(1-\alpha^*)R_2(P(\uh),P_R(\uh)).
\end{split}
\end{equation}

The Lagrangian can be written as
\begin{equation}
\begin{split}
\cL = & \alpha^* \mE_A\left[ \cC\left(2 P(\uh)\rho_1 |h_1|^2
\right)\right]+ \alpha^* \mE\left[ \cC\left(2
P_R(\uh)\rho_2|h_2|^2\right)\right] \\
& + \mE_{A^c}\left[\cC\left(2P(\uh)\rho_1 |h_1|^2\right)\right] \\
& + (1-\alpha^*)\mE_A \left[\cC\left( 2
P(\uh)\rho_3|h_3|^2\right)\right] \\
& -\lambda \Big(\mE[ P(\uh)]-P \Big) -\mu \Big(\mE[ P_R(\uh)]- P_R
\Big)
\end{split}
\end{equation}
where $\lambda$ and $\mu$ are Lagrange multipliers.

For $\uh \in A$, the KKT condition is given by:
\begin{equation}
\begin{split}
& \frac{\partial \cL}{\partial P(\uh)} \\
& \;\; = \frac{\alpha^*}{2\ln 2}\cdot \frac{1}{\frac{1}{2\rho_1
|h_1|^2}+ P(\uh)}+ \frac{1-\alpha^*}{2\ln 2}\cdot
\frac{1}{\frac{1}{2\rho_3 |h_3|^2}+
P(\uh)} \\
& \;\; \leq \lambda , \hspace{3.5cm} \text{with equality if }
P(\uh)
>0
\end{split}
\end{equation}
It is easy to check that $P^{(\alpha^*)}(\uh)$ for $\uh \in A$
given in \eqref{eq:fdp3} satisfies the preceding KKT condition.
The $P^{(\alpha^*)}(\uh)$ for $\uh \in A^c$ in \eqref{eq:fdp3} and
$P_R^{(\alpha^*)}(\uh)$ in \eqref{eq:fdpr3} also follow from the
KKT condition.

\section{Proof of Resource Allocation that Maximizes $C_{\mathtt{low}}$ \eqref{eq:fdlowII} for Scenario II}\label{app:scenarioII}

We let $R_1(\theta,P(\uh),P_R(\uh))$ and
$R_2(\theta,P(\uh),P_R(\uh))$ denote the two terms over which the
minimization in \eqref{eq:fdlowII} is taken. We can then express
\eqref{eq:fdlowII} in the following compact form:
\begin{equation}\label{eq:achieveII}
\begin{split}
C_{\mathtt{low}} & = \max\limits_{0\leq \theta \leq
1,(P(\uh),P_R(\uh)) \in \cG } \\
& \min \left \{ R_1(\theta,P(\uh),P_R(\uh)),\;
R_2(\theta,P(\uh),P_R(\uh)) \right \}.
\end{split}
\end{equation}

The max-min problem in \eqref{eq:achieveII} can be solved by using
Proposition \ref{prop:generule}. The main step is to obtain
$(\theta^{(\alpha)},P^{(\alpha)}(\uh),P^{(\alpha)}_R(\uh))$ that
maximizes
\begin{equation}
\begin{split}
& R(\alpha, \theta, P(\uh),P_R(\uh))\\
& \spp := \alpha
R_1(\theta,P(\uh),P_R(\uh))+(1-\alpha)R_2(\theta,P(\uh),P_R(\uh))
\end{split}
\end{equation}
for a given $\alpha$. The following lemma states that maximizing
the function $R(\alpha, \theta, P(\uh),P_R(\uh))$ over
$(\theta,P(\uh),P_R(\uh))$ is a convex programming problem and
hence can be solved by standard convex programming algorithms.
\begin{lemma}\label{lemma:convexII}
For a fixed $\alpha$, $R(\alpha, \theta, P(\uh),P_R(\uh))$ is a
concave function over $(\theta,P(\uh),P_R(\uh))$, where $0\leq
\theta \leq 1$ and $(P(\uh),P_R(\uh)) \in \cG$.
\end{lemma}

Lemma \ref{lemma:convexII} can be verified by computing the
Hessian of the function $R(\alpha, \theta, P(\uh),P_R(\uh))$ (for
a fixed $\alpha$) and showing that it is negative semidefinite.

We now apply Proposition \ref{prop:generule} to study the max-min
problem in \eqref{eq:achieveII} by considering the following three
cases.

\textbf{Case 1:} $\alpha^*=0$, and
$(\theta^{(0)},P^{(0)}(\uh),P^{(0)}_R(\uh))$ is an optimal
resource allocation, which needs to satisfy the condition
\begin{equation}\label{eq:fdcon1II}
R_1(\theta^{(0)},P^{(0)}(\uh),P^{(0)}_R(\uh)) \ge R_2
(\theta^{(0)},P^{(0)}(\uh),P^{(0)}_R(\uh)).
\end{equation}

We first derive $(\theta^{(0)},P^{(0)}(\uh),P^{(0)}_R(\uh))$ that
maximizes
\begin{equation}
R(0,\theta,P(\uh),P_R(\uh))=R_2(\theta,P(\uh),P_R(\uh)).
\end{equation}

It is easy to see that the optimal $\theta^{(0)}=1$ from the
expression of $R_2(\theta,P(\uh),P_R(\uh))$, and this results in
\begin{equation}\label{eq:fdcon1II2}
R_1(\theta^{(0)},P^{(0)}(\uh),P^{(0)}_R(\uh)) \leq R_2
(\theta^{(0)},P^{(0)}(\uh),P^{(0)}_R(\uh)).
\end{equation}
Comparing \eqref{eq:fdcon1II} and \eqref{eq:fdcon1II2}, it is
clear that only equality can be satisfied in \eqref{eq:fdcon1II}.
Hence this case can be included in the following case 3 with
$\alpha^*$ being allowed to take the value of $0$.

\textbf{Case 2:} $\alpha^*=1$, and
$(\theta^{(1)},P^{(1)}(\uh),P^{(1)}_R(\uh))$ is an optimal
resource allocation, which needs to satisfy the condition
\begin{equation}\label{eq:fdcon2II}
R_1(\theta^{(1)},P^{(1)}(\uh),P^{(1)}_R(\uh)) \leq R_2
(\theta^{(1)},P^{(1)}(\uh),P^{(1)}_R(\uh)).
\end{equation}

We first derive $(\theta^{(1)},P^{(1)}(\uh),P^{(1)}_R(\uh))$ that
maximizes
\begin{equation}
R(1,\theta,P(\uh),P_R(\uh))=R_1(\theta,P(\uh),P_R(\uh)).
\end{equation}

The Lagrangian can be written as
\begin{equation}
\begin{split}
\cL &= 2\theta \mE\left[\cC\left(\frac{P(\uh)\rho_1
|h_1|^2}{\theta} \right)\right] + 2\bartheta \mE\left[\cC\left(
\frac{P_R(\uh)\rho_2|h_2|^2}{\bartheta}\right)\right] \\
& \spp -\lambda \Big(\mE[ P(\uh)]-P \Big) -\mu \Big(\mE[
P_R(\uh)]- P_R \Big).
\end{split}
\end{equation}

It is easy to check that the KKT condition implies
\begin{equation}\label{eq:appfdp2II}
P(\uh)= \theta\left( \frac{1}{\lambda \ln 2}-\frac{1}{\rho_1
|h_1|^2} \right)^+
\end{equation}
\begin{equation}\label{eq:appfdpr2II}
P_R(\uh)= \bartheta\left( \frac{1}{\mu \ln
2}-\frac{1}{\rho_2|h_2|^2} \right)^+
\end{equation}
The KKT condition also implies that the optimal $\theta^{(1)}$
needs to satisfy the following condition:
\begin{equation}\label{eq:appfdtheta2II}
\begin{split}
& \frac{\partial \cL}{\partial \theta}= 2 \mE \left[
\cC\left(\frac{P(\uh)\rho_1 |h_1|^2}{\theta} \right)
\right]-2\mE\left[\cC\left(
\frac{P_R(\uh)\rho_2|h_2|^2}{\bartheta}\right)\right]
\\
& \; - \frac{1}{\ln 2}\mE \left[\frac{ P(\uh) \rho_1
|h_1|^2}{\theta+P(\uh) \rho_1 |h_1|^2}\right]+ \frac{1}{\ln
2}\mE\left[ \frac{ P_R(\uh) \rho_2
|h_2|^2}{\bartheta+P_R(\uh) \rho_2 |h_2|^2} \right] \\
& \;\; \begin{cases}
\leq 0, \pp \mbox{if }\; \theta = 0; \pp \text{(does not happen)} \\
=0, \pp \mbox{if }\; 0< \theta <1; \\
\ge 0, \pp \mbox{if }\; \theta =1. \pp \text{(does not happen)}
\end{cases}
\end{split}
\end{equation}
where the first and third cases do not happen because
$\frac{\partial \cL}{\partial \theta} \rightarrow \infty$  as
$\theta \rightarrow 0$, and $\frac{\partial \cL}{\partial \theta}
\rightarrow -\infty$ as $\theta \rightarrow 1$. It can also be
shown that $\frac{\partial \cL}{\partial \theta}$ is monotonically
decreasing for $0 \leq \theta \leq 1$. Hence $\frac{\partial
\cL}{\partial \theta}$ has at most one root for $0 \leq \theta
\leq 1$.

The iterative algorithm described in
\eqref{eq:fdp2II}-\eqref{eq:fdtheta2II} converges to the solution
of the KKT condition given in
\eqref{eq:appfdp2II}-\eqref{eq:appfdtheta2II}. Since the function
$R_1(\theta,P(\uh),P_R(\uh))$ is concave, the solution of the KKT
condition achieves the optimum. Condition \eqref{eq:compare2II}
follows from condition \eqref{eq:fdcon2II}.

\textbf{Case 3:} $0 < \alpha^* < 1$, and
$(\theta^{(\alpha^*)},P^{(\alpha^*)}(\uh),P^{(\alpha^*)}_R(\uh))$
is an optimal resource allocation, where $\alpha^*$ is determined
by the following condition
\begin{equation}\label{eq:fdcon3II}
\begin{split}
&
R_1(\theta^{(\alpha^*)},P^{(\alpha^*)}(\uh),P^{(\alpha^*)}_R(\uh))
\\
& \pp =
R_2(\theta^{(\alpha^*)},P^{(\alpha^*)}(\uh),P^{(\alpha^*)}_R(\uh)).
\end{split}
\end{equation}

We first derive
$(\theta^{(\alpha^*)},P^{(\alpha^*)}(\uh),P^{(\alpha^*)}_R(\uh))$
that maximizes
\begin{equation}
\begin{split}
& R(\alpha^*,\theta,P(\uh),P_R(\uh))\\
& \spp =\alpha^*
R_1(\theta,P(\uh),P_R(\uh))+(1-\alpha^*)R_2(\theta,P(\uh),P_R(\uh)).
\end{split}
\end{equation}
for given $\alpha^*$. The Lagrangian can be written as
\begin{equation}
\begin{split}
\cL =& 2\alpha^* \theta \mE_A \left[ \cC\left( \frac{P(\uh)\rho_1
|h_1|^2}{\theta} \right)\right] \\
& + 2\alpha^* \bartheta \mE \left[\cC\left(
\frac{ P_R(\uh)\rho_2|h_2|^2}{\bartheta}\right) \right] \\
& + 2\theta\mE_{A^c} \left[\cC\left(\frac{P(\uh)\rho_1
|h_1|^2}{\theta}\right)\right] \\
& + 2(1-\alpha^*)\theta \mE_A \left[ \cC\left( \frac{
P(\uh)\rho_3|h_3|^2}{\theta}\right) \right]\\
& -\lambda \Big(\mE[ P(\uh)]-P \Big) -\mu \Big(\mE[ P_R(\uh)]- P_R
\Big).
\end{split}
\end{equation}

For a given $\theta$, The optimal $(P(\uh),P_R(\uh))$ given in
\eqref{eq:fdp3II} and \eqref{eq:fdpr3II} follows from the KKT
condition. The KKT condition also implies that the optimal
$\theta$ for a given $(P(\uh),P_R(\uh))$ needs to satisfy the
following condition
\begin{equation} \label{eq:appfdtheta3II}
\begin{split}
& \frac{\partial \cL}{\partial \theta} \\
& = 2\alpha^* \mE_A\left[ \cC\left(\frac{P(\uh)\rho_1
|h_1|^2}{\theta} \right)\right] - \frac{\alpha^*}{\ln
2}\mE_A\left[ \frac{ P(\uh) \rho_1
|h_1|^2}{\theta+P(\uh) \rho_1 |h_1|^2} \right] \\
& -2\alpha^* \mE \left[\cC\left(
\frac{P_R(\uh)\rho_2|h_2|^2}{\bartheta}\right) \right]
+\frac{\alpha^*}{\ln 2}\mE \left[\frac{ P_R(\uh) \rho_2
|h_2|^2}{\bartheta+P_R(\uh) \rho_2|h_2|^2}\right]  \\
& + 2\mE_{A^c}\left[ \cC\left(\frac{P(\uh)\rho_1 |h_1|^2}{\theta}
\right)\right]- \frac{1}{\ln 2}\mE_{A^c} \left[\frac{ P(\uh)
\rho_1 |h_1|^2}{\theta+P(\uh) \rho_1 |h_1|^2}\right]
\\
& + 2(1-\alpha^*) \mE_A \left[\cC\left(\frac{P(\uh)\rho_3
|h_3|^2}{\theta} \right)\right] \\
& -\frac{1-\alpha^*}{\ln 2}\mE_A\left[ \frac{ P(\uh) \rho_3
|h_3|^2}{\theta+P(\uh) \rho_3 |h_3|^2}\right]  \\
& \begin{cases}
\leq 0, \pp \mbox{if }\; \theta = 0; \pp \text{(does not happen)} \\
=0, \pp \mbox{if }\; 0< \theta <1; \\
\ge 0, \pp \mbox{if }\; \theta =1. \pp \text{(does not happen)}
\end{cases}
\end{split}
\end{equation}
where the first and third cases do not happen because
$\frac{\partial \cL}{\partial \theta} \rightarrow \infty$ as
$\theta \rightarrow 0$, and $\frac{\partial \cL}{\partial \theta}
\rightarrow -\infty$ as $\theta \rightarrow 1$.

Therefore, the iterative algorithm described in
\eqref{eq:fdp3II}-\eqref{eq:fdtheta3II} converges to the solution
of the KKT condition. Since the function
$R(\alpha,\theta,P(\uh),P_R(\uh))$ is concave for a given
$\alpha$, the solution of the KKT condition achieves the optimum.
Condition \eqref{eq:compare3II} follows from condition
\eqref{eq:fdcon3II}.

\section{Proof of Resource Allocation that Maximizes $C_{\mathtt{low}}$ \eqref{eq:fdlowIII} for Scenario III}\label{app:scenarioIII}

We let $R_1(\theta(\uh),P(\uh),P_R(\uh))$ and
$R_2(\theta(\uh),P(\uh),P_R(\uh))$ denote the two terms over which
the minimization in \eqref{eq:fdlowIII} is taken. We can then
express \eqref{eq:fdlowIII} in the following compact form:
\begin{equation}\label{eq:achieveIII}
\begin{split}
C_{\mathtt{low}} & = \max\limits_{\begin{array}{l}\scriptstyle
0\leq \theta(\uh) \leq 1, \; \forall \; \uh \\ \scriptstyle
(P(\uh),P_R(\uh)) \in \cG
\end{array} } \\
& \min \Big \{ R_1(\theta(\uh),P(\uh),P_R(\uh)),\;
R_2(\theta(\uh),P(\uh),P_R(\uh)) \Big \}
\end{split}
\end{equation}

As in Scenario II, one main step to solve the max-min problem in
\eqref{eq:achieveIII} is to obtain
$(\theta^{(\alpha)}(\uh),P^{(\alpha)}(\uh),P^{(\alpha)}_R(\uh))$
that maximizes
\begin{equation}
\begin{split}
R(\alpha, \theta(\uh), P(\uh),P_R(\uh))& := \alpha
R_1(\theta(\uh),P(\uh),P_R(\uh)) \\
& +(1-\alpha)R_2(\theta(\uh),P(\uh),P_R(\uh))
\end{split}
\end{equation}
for a given $\alpha$. The following lemma states that maximizing
$R(\alpha, \theta(\uh), P(\uh),P_R(\uh))$ over
$(\theta(\uh),P(\uh),P_R(\uh))$ is a convex programming and hence
can be solved by standard convex programming algorithms.
\begin{lemma}\label{lemma:convexIII}
For a fixed $\alpha$, $R(\alpha, \theta(\uh), P(\uh),P_R(\uh))$ is
a concave function over $(\theta(\uh),P(\uh),P_R(\uh))$, where
$0\leq \theta(\uh) \leq 1$ for all $\uh$ and $(P(\uh),P_R(\uh))
\in \cG$.
\end{lemma}

Lemma \ref{lemma:convexIII} can be verified by computing the
Hessian of the function $R(\alpha, \theta(\uh), P(\uh),P_R(\uh))$
(for a fixed $\alpha$) and showing that it is negative
semidefinite.

As for Scenarios I and II, we apply Proposition
\ref{prop:generule} to study the max-min problem in
\eqref{eq:achieveIII} by considering the following three cases.

\textbf{Case 1:} $\alpha^*=0$, and
$(\theta^{(0)}(\uh),P^{(0)}(\uh),P^{(0)}_R(\uh))$ is an optimal
resource allocation, which needs to satisfy the condition
\begin{equation}\label{eq:fdcon1III}
\begin{split}
& R_1(\theta^{(0)}(\uh),P^{(0)}(\uh),P^{(0)}_R(\uh)) \\
& \pp \ge R_2 (\theta^{(0)}(\uh),P^{(0)}(\uh),P^{(0)}_R(\uh)).
\end{split}
\end{equation}

We first derive $(\theta^{(0)}(\uh),P^{(0)}(\uh),P^{(0)}_R(\uh))$
that maximizes
\begin{equation}
R(0,\theta(\uh),P(\uh),P_R(\uh))=R_2(\theta(\uh),P(\uh),P_R(\uh)).
\end{equation}

It is clear that $\theta^{(0)}(\uh)$ given in
\eqref{eq:fdtheta1III} is optimal from the expression of
$R_2(\theta(\uh),P(\uh),P_R(\uh))$. The power allocation
$P^{(0)}(\uh)$ given in \eqref{eq:fdp1III} then easily follows
from the KKT condition.

For case 1 to happen, condition \eqref{eq:fdcon1III} needs to be
satisfied. To characterize the most general condition for case 1
to happen, for the given $\theta^{(0)}(\uh)$ and $P^{(0)}(\uh)$
$P_R(\uh)$ needs to maximize
$R_1(\theta^{(0)}(\uh),P^{(0)}(\uh),P_R(\uh))$, which has the
following form:
\begin{equation}\label{eq:r1case1III}
\begin{split}
& R_1(\theta^{(0)}(\uh),P^{(0)}(\uh),P_R(\uh)) \\
& \pp = 2\mE_{\{\uh: P^{(0)}(\uh)=0\}} \left[\cC\left(
P_R(\uh)\rho_2|h_2|^2 \right)\right]
\end{split}
\end{equation}
The optimal $P_R^{(0)}(\uh)$ given in \eqref{eq:fdpr1III} then
follows from the KKT condition. Finally, condition
\eqref{eq:compare1III} follows from condition
\eqref{eq:fdcon1III}.

The proofs for cases 2 and 3 are similar to those for Scenario II
given in Appendix \ref{app:scenarioII}, and are omitted.
\section*{Acknowledgment}
The authors would like to thank Dr.\ Gerhard Kramer of Bell
Laboratories, Alcatel-Lucent, for helpful discussions on the topic
of parallel relay channels. The authors would also like to thank
Dr.\ Jianwei Huang of Princeton University for helpful discussions
on the topic of optimization.

\bibliographystyle{IEEEtran}
%\bibliography{relay}

\begin{biographynophoto}{Yingbin Liang}(S'01--M'05)
received the Ph.D.\ degree in Electrical Engineering from the
University of Illinois at Urbana-Champaign in 2005. Since
September 2005, she has been working as a postdoctoral research
associate at Princeton University. Her research interests include
information theory, wireless communications, and wireless
networks.

Dr. Liang was a Vodafone Fellow at the University of Illinois at
Urbana-Champaign during 2003-2005, and
received the Vodafone-U.S. Foundation Fellows Initiative Research
Merit Award in 2005. She also received the M. E. Van Valkenburg
Graduate Research Award from the ECE department, University of
Illinois at Urbana-Champaign, in 2005.
\end{biographynophoto}

\begin{biographynophoto}{Venugopal V. Veeravalli}(S'86--M'92--SM'98--F'06)
received the Ph.D.\ degree in 1992 from the University of Illinois
at Urbana-Champaign, the M.S.\ degree in 1987 from Carnegie-Mellon
University, Pittsburgh, PA, and the B.\ Tech.\ degree in 1985 from
the Indian Institute of Technology, Bombay, (Silver Medal Honors),
all in Electrical Engineering. He joined the University of
Illinois at Urbana-Champaign in 2000, where he is currently a
Professor in the department of Electrical and Computer
Engineering, and a Research Professor in the Coordinated Science
Laboratory. He served as a program director for communications
research at the U.S. National Science Foundation in Arlington, VA
from 2003 to 2005. He was an assistant professor at Cornell
University, Ithaca, NY from 1996 to 2000.

His research interests include distributed sensor systems and
networks, wireless communications, detection and estimation
theory, and information theory. He is a Fellow of the IEEE and
currently on the Board of Governors of the IEEE Information Theory
Society. He was an Associate Editor for Detection and Estimation
for the IEEE Transactions on Information Theory from 2000 to 2003,
and an associate editor for the IEEE Transactions on Wireless
Communications from 1999 to 2000. Among the awards he has received
for research and teaching are the IEEE Browder J. Thompson Best
Paper Award in 1996, the National Science Foundation CAREER Award
in 1998, and the Presidential Early Career Award for Scientists
and Engineers (PECASE) in 1999.
\end{biographynophoto}

\begin{biographynophoto}{H. Vincent Poor}(S'72--M'77--SM'82--F'87)
received the Ph.D. degree in EECS from Princeton University in
1977. From 1977 until 1990, he was on the faculty of the
University of Illinois at Urbana-Champaign. Since 1990 he has been
on the faculty at Princeton, where he is the Dean of Engineering
and Applied Science, and the Michael Henry Strater University
Professor of Electrical Engineering. Dr. Poor's research interests
are in the areas of stochastic analysis, statistical signal
processing and their applications in wireless networks and related
fields. Among his publications in these areas is the recent book
{\em MIMO Wireless Communications} (Cambridge University Press,
2007).

Dr. Poor is a member of the National Academy of Engineering, a
Fellow of the American Academy of Arts and Sciences, and a former
Guggenheim Fellow. He is also a Fellow of the Institute of
Mathematical Statistics, the Optical Society of America, and other
organizations. In 1990, he served as President of the IEEE
Information Theory Society, and in 2004-07 as the Editor-in-Chief
of these {\em Transactions}. Recent recognition of his work
includes the 2005 IEEE Education Medal and the 2007 IEEE Marconi
Prize Paper Award.
\end{biographynophoto}


%\vfill


\end{document}